\documentclass[aps,prl,twocolumn,epsfig,pdflatex]{revtex4-1}

\usepackage{amsmath,amssymb}
\usepackage{graphicx}

\usepackage{float}
\usepackage[usenames,dvipsnames]{xcolor}
\usepackage{amsthm,comment}
\usepackage{booktabs}
\usepackage[export]{adjustbox}
\usepackage[section]{placeins}
\usepackage[caption=false]{subfig}
\usepackage[percent]{overpic}
\bibpunct{[}{]}{,}{n}{}{}
\definecolor{red1}{HTML}{FF4136}
\definecolor{green1}{HTML}{00802b}

\usepackage[hidelinks]{hyperref} 

\hypersetup{colorlinks=true,citecolor=Red,linkcolor=Blue,urlcolor=Blue}
\usepackage[export]{adjustbox}

\graphicspath{{Figures/}}

\begin{document}

\title {Origin of the Magnetic and Orbital ordering in $\alpha$-Sr$_2$CrO$_4$}

\author{Bradraj Pandey$^{1,2}$, Yang Zhang$^{1}$, Nitin Kaushal$^{1,2}$,
Rahul Soni$^{1,2}$, Ling-Fang Lin$^{1}$, Wen-Jun Hu$^{1}$, Gonzalo Alvarez$^{3}$,  
and Elbio Dagotto$^{1,3}$}
\affiliation{$^1$Department of Physics and Astronomy, University of Tennessee, Knoxville, Tennessee 37996, USA \\ 
$^2$Materials Science and Technology Division, Oak Ridge National Laboratory, Oak Ridge, Tennessee 37831, USA \\ 
$^3$Computational Sciences $\&$ Engineering Division and Center for Nanophase Materials Sciences,
Oak Ridge National Laboratory,~Oak Ridge,~Tennessee 37831,~USA
 }

\begin{abstract}

Motivated by recent experimental progress in transition metal oxides
with the K$_2$NiF$_4$ structure,  we investigate the magnetic and orbital ordering 
in $\alpha$-Sr$_2$CrO$_4$.
Using first principles calculations, first we derive a three-orbital Hubbard model,
which reproduces the {\it ab initio} band structure near the Fermi level.
The unique reverse splitting of $t_{2g}$ orbitals in $\alpha$-Sr$_2$CrO$_4$,
 with the $3d^2$ electronic configuration for the Cr$^{4+}$ oxidation state,
opens up the possibility of orbital ordering in this material.
Using real-space Hartree-Fock for multi-orbital systems, we constructed the ground
state phase diagram for the two dimensional compound $\alpha$-Sr$_2$CrO$_4$.
We found stable ferromagnetic, antiferromagnetic, antiferro-orbital, and
 staggered orbital stripe ordering in robust regions of the phase diagram.
Furthermore, using the density matrix renormalization group method for two-leg ladders 
with the realistic hopping parameters of $\alpha$-Sr$_2$CrO$_4$, we explore
 magnetic and orbital ordering for experimentally relevant interaction
 parameters. Again, we find a clear signature of antiferromagnetic spin ordering
 along with antiferro-orbital ordering at moderate to large
 Hubbard interaction strength. We also explore the orbital-resolved
density of states with Lanczos, predicting insulating behavior for the
compound $\alpha$-Sr$_2$CrO$_4$, in agreement with experiments. Finally, an intuitive understanding
of the results is provided based on a hierarchy between orbitals, with $d_{xy}$ driving the spin 
order, while electronic repulsion and the effective one dimensionality of the movement within the
$d_{xz}$ and $d_{yz}$  orbitals driving the orbital order.

\end{abstract}


\maketitle

\section{I. Introduction}

Transition-metal oxides with the perovskite structure exhibit a wide variety of exotic magnetic, charge, and orbital ordering~\cite{kugel1,kugel,dagotto2001}.
The rich phase diagrams and intriguing physical properties of these materials is due to the 
Hubbard and Hund interactions
among the electrons occupying the $3d$ orbitals~\cite{johnston2010,pdai,elbio,chubukov}.
In particular, the study of perovskite compounds with the K$_2$NiF$_4$ structure is of considerable 
interest due to their similar crystal structures  to the widely studied
high-$T_c$ cuprates based on La$_2$CuO$_4$~\cite{scalapino,dagotto94}
and also the exotic $p$-wave superconductor Sr$_2$RuO$_4$~\cite{Maeno}.
The orbital degree of freedom plays a crucial role in various
types of structural transitions~\cite{San}, in magnetic and charge order~\cite{ezhov},
and in exotic phenomena, such as the colossal magnetoresistance in transition-metal oxides 
with perovskite structure~\cite{tokura} and the previously mentioned high temperature superconductivity.
The observation of the peculiar antiferromagnetism in metallic
transition metal oxides~\cite{komarek} and ferromagnetism in  insulating
transition metal oxides~\cite{Jun,Varignon} are often caused by the phenomenon of orbital
ordering in the system~\cite{khoms}.

Recent developments in the chromium-based Ruddlesden-Popper (RP) 
series Sr$_{n+1}$Cr$_n$O$_{3n+1}$,  provide an ideal playground for the spin and orbital
degrees of freedom. Using X-rays and neutron diffraction, varying temperature,
the simultaneous development of orbital and  magnetic ordering has been observed 
for Sr$_3$Cr$_2$O$_7$, the $n=2$ member of the RP series~\cite{Jean}.
In this compound, the  spin ordering was found to be  antiferromagnetic, while the orbital ordering
was described as forming orbital singlet states~\cite{Jean,helman}. The possibility of high-$T_c$ superconductivity in Sr$_3$Cr$_2$O$_7$
has been also proposed due to the hidden-ladder electronic structures ~\cite{ogura} present in this compound. The perovskite SrCrO$_3$ with a cubic structure
(the $n=\infty$ member of the RP system) was synthesized five decades ago and it is believed to be a non-magnetic metal~\cite{chamber}. 
More recent studies on poly-crystalline SrCrO$_3$ samples
under high pressure demonstrated an anomalous non-metallic behavior~\cite{Zhou}. Based on neutron and powder X-ray diffraction,
orbital ordering and electronic phase coexistence (tetragonal and cubic phases) was observed in SrCrO$_3$~\cite{San}. 
At $T=40$~K  due to the orbital ordering instability, the cubic structure  transforms to an antiferromagnetic 
tetragonal phase, which results on a low-temperature phase coexistence in SrCrO$_3$~\cite{San}. 
Interestingly, orbital-ordering induced ferroelectricity has been proposed in SrCrO$_3$~\cite{Gupta}.

Recently the study of $\alpha$-Sr$_2$CrO$_4$  (the $n=1$ member of the RP series. with $\alpha$ denoting the allotrope with layered structure) received attention due to its exotic magnetic and orbital ordering~\cite{sakurai,weng,nozaku}. 
This compound has the K$_2$NiF$_4$ type structure, rendering it isostructural to high-$T_c$ superconducting cuprates such as La$_2$CuO$_4$.
In Sr$_2$CrO$_4$, chromium is in a Cr$^{4+}$ oxidation state 
with a $3d^2$ electronic configuration and shows rare and unusual reversed
crystal-field splitting~\cite{takashi}. Although the  compound $\alpha$-Sr$_2$CrO$_4$ 
was first synthesized long time ago, high-quality bulk samples were produced only quite recently~\cite{Hsakurai}.
In a recent experiment~\cite{Hsakurai} on pure samples
of $\alpha$-Sr$_2$CrO$_4$, using magnetic susceptibility and specific heat measurements,
two successive phase transitions at $T_N=112$~K and $T_S=140$~K were reported. As discussed in Ref.~\cite{takashi}, the lower temperature phase transition ($T_N=112$~K) is attributed to N\'eel ordering, while the higher temperature transition $T_S=140$~K is caused by orbital ordering~\cite{takashi}.

\begin{figure} [H]
\centering
\includegraphics[width=0.48\textwidth]{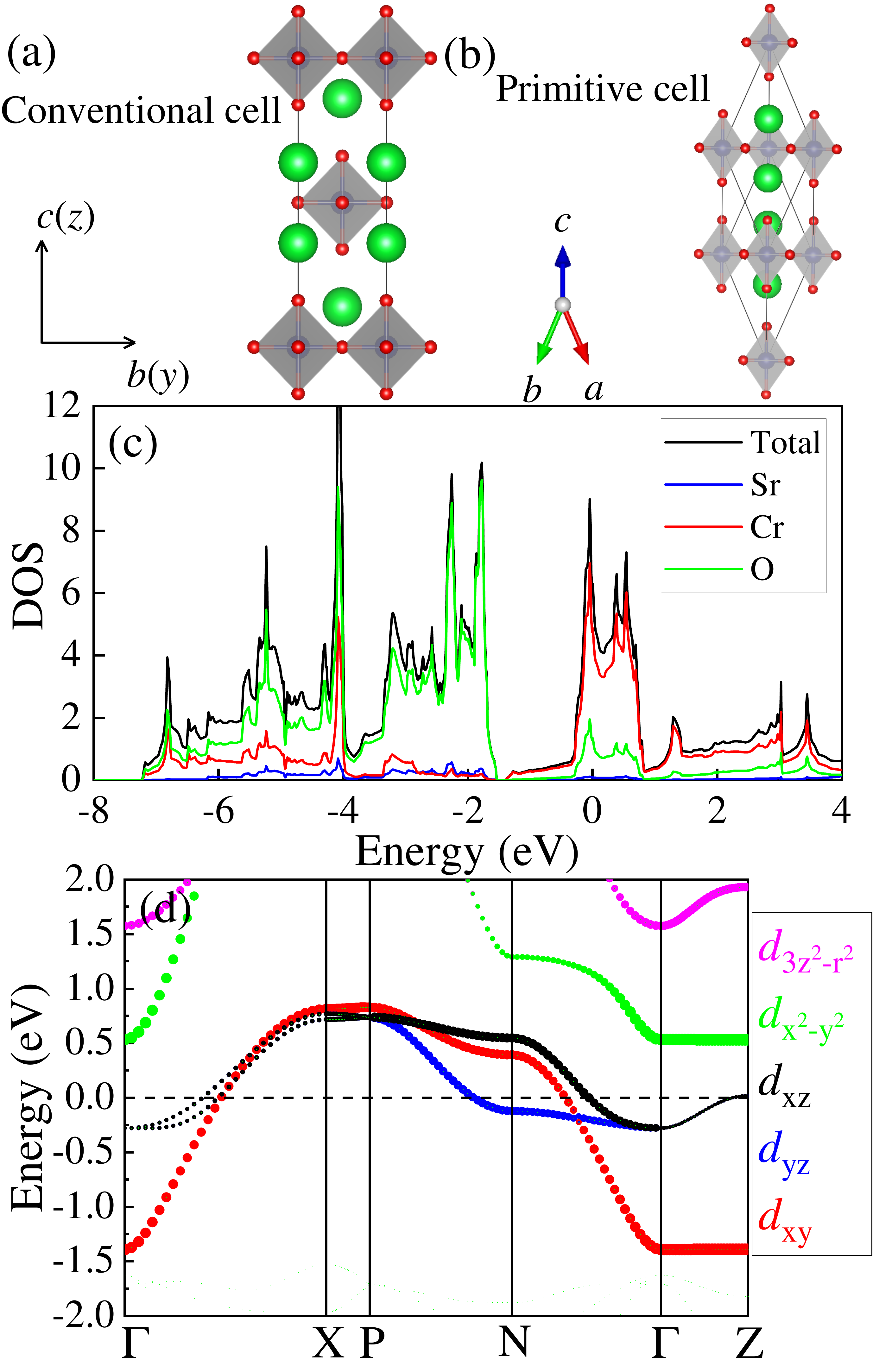}
\caption{(a) Schematic crystal structure of the canonical cell of Sr$_2$CrO$_4$ with the convention: Green = Sr; Blue = Cr; Red = O. (b) Schematic crystal structure of the primitive cell of Sr$_2$CrO$_4$. (b) Density of states near the Fermi level for the non-magnetic state. Black = Total; Blue = Sr; Red = Cr; Green = O. (c) Projected band structure of Sr$_2$CrO$_4$ for the non-magnetic state. The Fermi level is shown with dashed lines. The weight of each chromium orbital is represented by the size of the circle. The Brillouin zone notation is $\Gamma=(0,0,0)$, 
${\rm X}=(0,0,\pi/2)$, ${\rm X}=(0,0,\pi/2)$, ${\rm P}=(\pi/4,\pi/4,\pi/4)$, ${\rm N}=(0,\pi/2,0)$, and ${\rm Z}=(\pi/2,\pi/2,-\pi/2)$.}
\label{Fig1}
\end{figure}

In Refs.~\cite{takashi,justin} using density functional theory, the orbital ordering in $\alpha$-Sr$_2$CrO$_4$  was explained by the reversal of the crystal-field splitting.
More specifically, it was shown that the 
 crystal-field energy location of the $3d_{xy}$ orbital of the chromium ion is 
lower in energy compared to the doubly degenerate   $3d_{xz}$ and $3d_{yz}$ orbitals, which leads to an active orbital degree of freedom in the system. 
Moreover, in another experiment~\cite{sakurai}, the pressure(P)-temperature(T) phase diagram 
was obtained for $\alpha$-Sr$_2$CrO$_4$, showing that this material remains an insulator even at large pressure 
and temperature. Interestingly, under the high pressure condition they  observed 
only one phase transition from the antiferromagnetic insulating phase to a high temperature paramagnetic phase, while the orbital-ordering phase transition ($T_S$) disappears~\cite{sakurai}. The disappearance of orbital-ordering was explained by the restoration of the reversed crystal-field 
splitting under high-pressure~\cite{yamaguchi}. 
This shows the importance of the rare reverse splitting in the orbitally ordered 
compound $\alpha$-Sr$_2$CrO$_4$. 
Using resonant X-ray scattering collinear N\'eel-type magnetic ordering coexisting with stripe-like ordering was demonstrated in an experimental study of $\alpha$-Sr$_2$CrO$_4$~\cite{zhu}. Due to the difficulties in the 
synthesis of $\alpha$-Sr$_2$CrO$_4$ in pure form  and due to the effect of strong electronic interactions, 
only a few experimental and theoretical studies addressing this material have been presented. 

As discussed above, previous theoretical studies for this material were based mainly on
the density functional theory. In this publication, for the first time, 
we have studied the magnetic and orbital ordering of $\alpha$-Sr$_2$CrO$_4$ using a multiorbital Hubbard Hamiltonian incorporating the Hubbard and Hund interactions. Via first principles calculations we obtain the hopping amplitudes 
for the two-dimensional compound $\alpha$-Sr$_2$CrO$_4$. 
Employing the unrestricted real-space Hartree-Fock approximation for 
two-dimensional three-orbital Hubbard model, we constructed the ground state 
phase diagram by varying the on-site Hubbard repulsion $U$ and Hund coupling $J_H$. 
We have found interesting spin- and orbital-ordered states in the phase diagram, including ferromagnetic, antiferromagnetic, antiferro-orbital, and staggered orbital stripe ordering, varying the $U$ and $J_H$ parameters. 
More importantly, we find a robust insulating phase with 
antiferromagnetic spin ordering and antiferro-orbital ordering
in a large region of the phase diagram, which we consider to be quite relevant for 
the experimental study of the compound $\alpha$-Sr$_2$CrO$_4$.
Furthermore, employing the density matrix renormalization group (DMRG) method~\cite{white}
for a two leg-ladder with the realistic hopping parameters of $\alpha$-Sr$_2$CrO$_4$, 
we explore the spin and orbital ordering at a fixed Hund coupling $J_{H}/U=0.2$~\cite{hundiron}. 
Interestingly, we found an excellent agreement between Hartree-Fock 
and the DMRG method with regards to spin and orbital ordering for 
experimentally relevant interaction parameters.
 Using DMRG, we find the same insulating state with
 antiferromagnetic spin ordering and antiferro-orbital ordering 
 as found via Hartree-Fock, for interaction strength $U/W \gtrsim 2.0$.
 We have also obtained the orbital-resolved density of states using the Lanczos method~\cite{dagotto94},
 which predicts an insulating ground state for $\alpha$-Sr$_2$CrO$_4$. In a recent
 experiment, the insulating nature of the ground state with antiferromagnetic spin order
 was demonstrated for $\alpha$-Sr$_2$CrO$_4$.

The organization of the manuscript is as follows. Section II provides details of the {\it ab initio} calculations for $\alpha$-Sr$_2$CrO$_4$. 
Section III contains the multiorbital model and details of the numerical methods used. Section IV presents
the results of the real-space Hartree-Fock method, 
where an extended phase diagram of the model was constructed. 
Section V has the DMRG and Lanczos results, 
where we focus on Hund coupling $J_H/U=0.2$. In Section VI, a simple rationalization for the results 
we have found is provided, explaining both the magnetic and orbital order based on electronic correlations.
Finally, in Section VII we present our conclusions.  

\section{II. DFT Methods}
Under ambient conditions, $\alpha$-Sr$_2$CrO$_4$ forms a quasi-two-dimensional K$_2$NiF$_4$-type structure with the space group $I4/mmm$ (No. 139), as shown in Fig.~\ref{Fig1}(a). The experimental lattice parameters are $a=b=3.816$ \AA ~and $c=12.482$~\AA ~\cite{sakurai}. To understand the electronic properties of the $\alpha$-Sr$_2$CrO$_4$ system, first-principles density functional theory (DFT) calculations were performed based on the projector augmented wave (PAW) method, as implemented in the Vienna {\it ab initio} simulation package (VASP) code~\cite{Kresse:Prb,Kresse:Prb96,Blochl:Prb}. Here, we calculated the electronic correlations by using the generalized gradient approximation (GGA) with the Perdew-Burke-Ernzerhof  pseudopotentials \cite{Perdew:Prl}.

For the non-magnetic state, our calculation uses the primitive cell instead of the conventional cell to 
evaluate the electronic structure of Sr$_2$CrO$_4$, with the volume of the primitive cell being half of the conventional cell.
Figure~\ref{Fig1}(b) shows the primitive lattice vectors $a_1$=($-a/2$, $a/2$, $c/2$),
$a_2$=($a/2$, $-a/2$, $c/2$), and $a_3$=($a/2$, $a/2$, $-c/2$), where $a$ and $c$ are the conventional-cell lattice constants.
The plane-wave cutoff energy was $600$~eV and the adopted $k$-point mesh was $10\times10\times10$. 
Note that we tested explicitly that this $k$-point mesh already leads to converged results. 
In addition to the standard DFT calculation discussed thus far, the maximally localized Wannier functions 
(MLWFs) method was employed to study the three Cr $3d$ $t_{\rm 2g}$ bands by using the WANNIER90 packages~\cite{Mostofi:cpc}.

Furthermore, we also followed the local spin density approach (LSDA) plus $U$, within the Dudarev formulation~\cite{Dudarev:prb} in the magnetic DFT calculations. Since no significant structural transition was reported at low temperatures by experiments~\cite{zhu,chuang}, 
we have used the same crystal structure for the magnetic states
as employed for the non-magnetic calculations. To better understand and focus on the electronic 
correlations, we did not relax the lattice constant and atomic position for the
magnetic configurations that we studied. The magnetic lattice was chosen 
as a $\sqrt{2}\times\sqrt{2}\times1$ supercell, involving two Cr atoms in one plane 
with the lattice constants $5.397$~\AA ~and $c=12.482$~\AA, respectively. 
In our density of states (DOS) magnetic calculations, 
we used $12\times12\times8$ $k$-points and the plane cutoff energy was $550$~eV.

\begin{figure}
\centering
\includegraphics[width=0.48\textwidth]{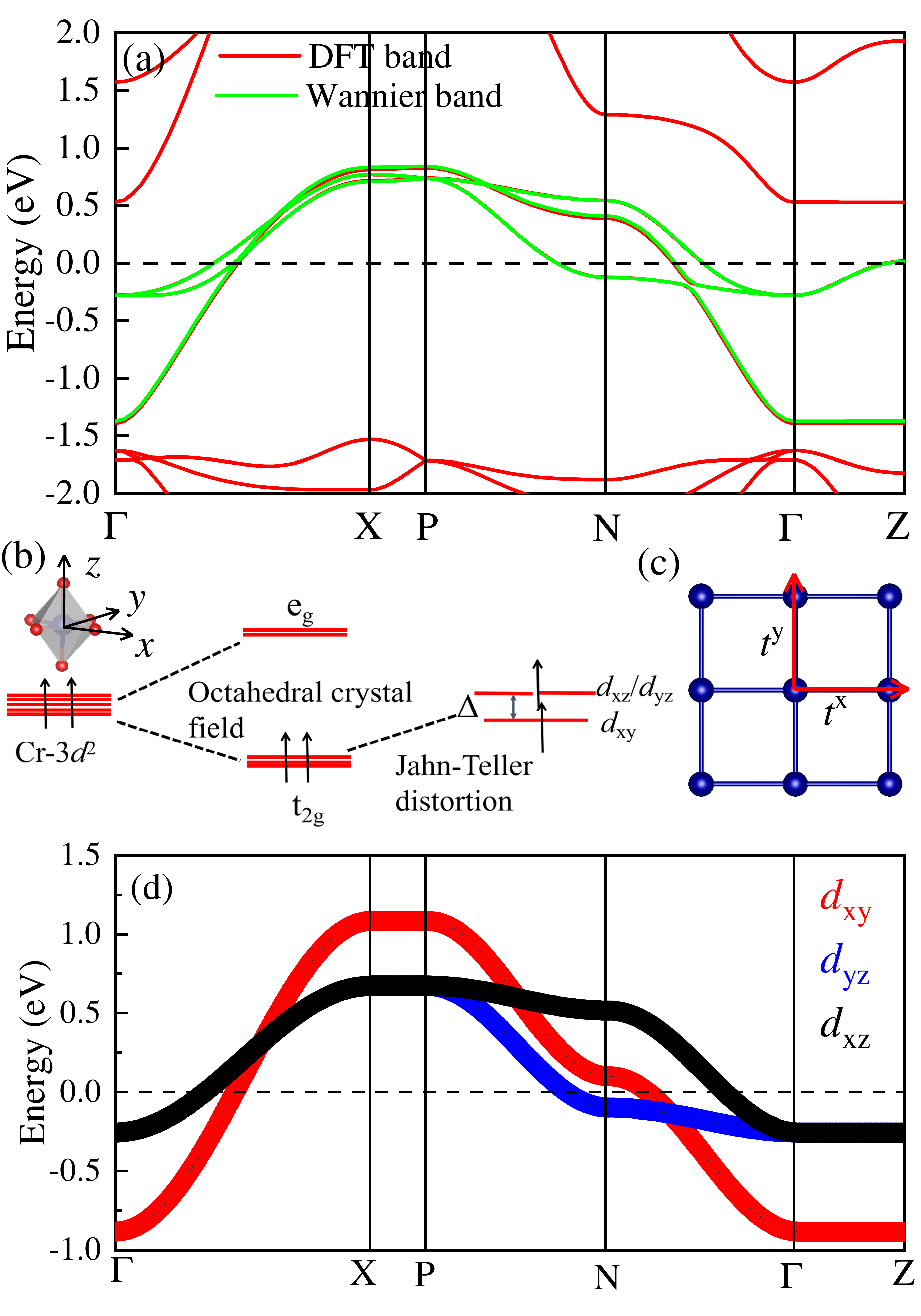}
\caption{(a) The original DFT band dispersion for Sr$_2$CrO$_4$ is shown using red solid lines, while the Wannier interpolated band dispersion is presented using green dashed lines. (b) Schematic energy splitting of Cr's $3d$ orbitals with the $d^2$ configuration. (c) Sketch of nearest-neighbor hoppings along the $x$ and $y$ directions, as indicated. (d) Tight binding (TB) band structure for the three $t_{\rm 2g}$ orbitals using the 3$\times$3 nearest-neighbor hopping matrices described in Sec.~III. Note that along $\Gamma$-P-X, the $d_{yz}$ and $d_{xz}$ are identical.}
\label{Fig2}
\end{figure}

Let us discuss now the electronic structure corresponding to the non-magnetic (NM) state of Sr$_2$CrO$_4$. According to the calculated DOS [see Figs.~\ref{Fig1}(b-c)], the bands near the Fermi level are primarily contributed by the Cr-$3d$ $t_{2g}$ orbitals, slightly $hybridized$ with the O-$2p$ orbitals.
Figure~\ref{Fig1}(c) shows that the $e_g$ orbitals $d_{x^2-y^2}$ and $d_{3z^2-r^2}$ occupy high-energy states in the conduction band, 
indicating this system can be accurately regarded as having two electrons per site on the three $t_{\rm 2g}$ orbitals. 
For this reason, we constructed three Wannier functions based on the MLWFs method~\cite{Mostofi:cpc}, 
involving the $t_{2g}$ orbital basis $d_{xy}$, $d_{yz}$, and $d_{xz}$ for each Cr atom.
As shown in Fig.~\ref{Fig2}(a), the DFT bands are  accurately reproduced by the Wannier bands obtained from MLWFs. 
Based on the information of Wannier functions, we can deduce the on-site energy of the three $t_{2g}$ orbitals
and the corresponding hopping parameters (see Sec.~III for details).

The energy splitting of the Cr $3d$ orbitals is sketched in Fig.~\ref{Fig2}(b). First, the octahedral crystal field leads to three lower-energy $t_{2g}$ orbitals ($d_{xy}$, $d_{yz}$, and $d_{xz}$) and two higher-energy $e_g$ orbitals ($d_{x^2-y^2}$ and $d_{3z^2-r^2}$). In addition, the Jahn-Teller distortion produces two different types of Cr-O bonds, with two longer Cr-O bonds along the $z$ direction and four shorter Cr-O bonds within the $a-b$ plane, resulting in the energy of the $d_{xy}$ orbital shifted down compared with the energies of the $d_{yz}$ and $d_{xz}$ orbitals. Thus, this system can be regarded as ($d_{xy}$)$^1$($d_{xz}$,$d_{yz}$)$^1$, as illustrated in Fig.~\ref{Fig2}(b). Based on the on-site energy difference between the $d_{xy}$ and $d_{xz}$/$d_{yz}$ orbitals, the crystal splitting energy is $\Delta = 0.11$ eV. Because it is too difficult to deal with hopping matrices over extended distances when employing three orbitals in DMRG, we only considered the nearest neighbor (NN) hopping matrices along the $x$ and $y$ axes [Fig.~\ref{Fig2}(c)].
Figure~\ref{Fig2}(d) shows that the tight binding (TB) band structure for three $t_ {2g}$ orbitals using 
\emph{only} the NN hopping matrix qualitatively agrees with the DFT band structure.

\section{III. Three-Orbital Hubbard Model and Methods}

The multiorbital Hubbard model for the  primarily two-dimensional compound Sr$_2$CrO$_4$ 
with three Cr orbitals at each site, derived using the {\it ab initio} calculation of the previous section, will be presented here in detail. This multiorbital Hubbard model can be written as the sum 
of kinetic and interaction energy terms $H=H_k+H_{in}$~\cite{Luo}. 
The kinetic component contains the hopping along the $x$-direction and $y$-direction
 of the two-dimensional lattice:
\begin{eqnarray}
H_k = \sum_{i,\sigma,\gamma,\gamma'} t^{x}_{\gamma, \gamma'}\left(c^{\dagger}_{i\sigma,\gamma}c^{\phantom\dagger}_{i+{\hat x}, \sigma, \gamma'}+H.c.\right) \nonumber \\
+  t^{y}_{\gamma, \gamma'}\left(c^{\dagger}_{i,\sigma,\gamma}c^{\phantom\dagger}_{i+{\hat y}, \sigma, \gamma'}+H.c.\right)+ \sum_{i,\gamma \sigma} \Delta_{\gamma} n_{i,\sigma, \gamma},
\end{eqnarray}
where $t^{x}_{\gamma, \gamma'}$ is the NN hopping matrix  
along the $x$-direction in the orbital space $\gamma=\{d_{xz},d_{yz},d_{xy}\}$, while 
$t^{y}_{\gamma, \gamma'}$ is the NN hopping matrix along the $y$-direction. The vectors ${\hat x}$ and ${\hat y}$
are unit vectors (in lattice spacing units) along the $x$ and $y$ axes, respectively.
$n_{i,\sigma, \gamma}$ represents the orbital- and spin-resolved electronic number operator. 
These three orbitals will be denoted  as $\gamma = \{1,2,3\}$, respectively, for notation simplicity. 
The hopping matrices for $\alpha$-Sr$_2$CrO$_4$ were obtained from a 
tight-binding Wannier function analysis of DFT results and they are all in eV units.  
The 3$\times$3 hopping matrix along the $x$-direction $t^{x}_{\gamma, \gamma'}$, between sites $i$ and $i+{\hat x}$, in orbital space and in eV units, is given by:
\[
t^{x}_{\gamma, \gamma'}=
  \begin{bmatrix}
    -0.193 & 0.000 & 0.000 \\
     0.000 & -0.039 & 0.000  \\
     0.000 & 0.000 & -0.246 
  \end{bmatrix}
\] 
where $\gamma$ is the orbital index for site $i$ and  $\gamma'$ for $i+{\hat x}$. Similarly,
$t^{y}_{\gamma, \gamma'}$ is the 3$\times$3 hopping matrix between sites $i$ and $i+{\hat y}$ along the $y$-direction:
\[
t^{y}_{\gamma, \gamma'}=
  \begin{bmatrix}
    -0.039 & 0.000 & 0.000 \\
     0.000 & -0.193 & 0.000  \\
     0.000 &  0.000 & -0.246 
  \end{bmatrix}
\]
The on-site matrix with the crystal fields $\Delta_{\gamma}$ for each orbital is given by:
\[
t^{OnSite}_{\gamma, \gamma}=
  \begin{bmatrix}
     4.748 & 0.000 & 0.000 \\
     0.000 & 4.748 & 0.000  \\
     0.000 &  0.000 & 4.639 
  \end{bmatrix}
\]
The kinetic energy bandwidth is $W=2.0$~eV.

The electronic interaction portion of the Hamiltonian is:
\begin{eqnarray}
H_{in}= U\sum_{i\gamma}n_{i\uparrow \gamma} n_{i\downarrow \gamma} +\left(U'-\frac{J_H}{2}\right) \sum_{i,\gamma < \gamma'} n_{i \gamma} n_{i\gamma'} \nonumber \\
-2J_H  \sum_{i,\gamma < \gamma'} {{\bf S}_{i,\gamma}}\cdot{{\bf S}_{i,\gamma'}}+J_H  \sum_{i,\gamma < \gamma'} \left(P^+_{i\gamma} P_{i\gamma'}+H.c.\right). 
\end{eqnarray}
The first term is the on-site Hubbard repulsion between $\uparrow$ and $\downarrow$ electrons in the same orbital. 
The second term is the on-site electronic repulsion between electrons at 
different orbitals, same site. Due to the $SU(2)$ symmetry of the Hamiltonian, the standard relation $U'=U-2J_H$ is here assumed. 
The third term shows the ferromagnetic Hund's interaction between electrons occupying the active three orbitals $(\gamma=\{d_{xz},d_{yz},d_{xy}\})$. 
The operator ${\bf S}_{i,\gamma}$ is the total spin at site $i$ and orbital $\gamma$. The last term is the on-site
pair-hopping between different orbitals, where $P_{i \gamma}$=$c_{i \downarrow \gamma} c_{i \uparrow \gamma}$. 

To solve this three-orbital Hubbard model, and obtain the predicted ground state properties of
$\alpha$-Sr$_2$CrO$_4$, three many-body techniques will be employed: the real-space Hartree-Fock, DMRG,
and Lanczos methods. For the real-space Hartree-Fock calculation 
we used a cluster size up to $ 16 \times 16$,
while for DMRG we used cluster sizes up to  $2 \times 10$ (sizes are severely restricted within DMRG three-orbitals
because, due to entanglement, this cluster demands even more effort than a $6 \times 10$ one-orbital). 
Using the Hartree-Fock method, we have calculated the density of state (DOS), spin and orbital correlations and
their structure factors.
The electronic density was fixed at $n=2/3$ (two electrons per site, i.e. two electrons in three orbitals) in our numerical calculations.
For both the real-space Hartree-Fock  and DMRG methods we employed open-boundary conditions.
For DMRG, at least 1600 states were kept during the calculations and we used the DMRG++ software~\cite{gonzalo}. Furthermore, we employed the Lanczos method for small clusters $L=2 \times 2 $ 
to obtain the orbital-resolved density of states. 

\section{IV. Hartree-Fock Results}
\begin{figure}[h]
\centering
\rotatebox{0}{\includegraphics*[width=\linewidth]{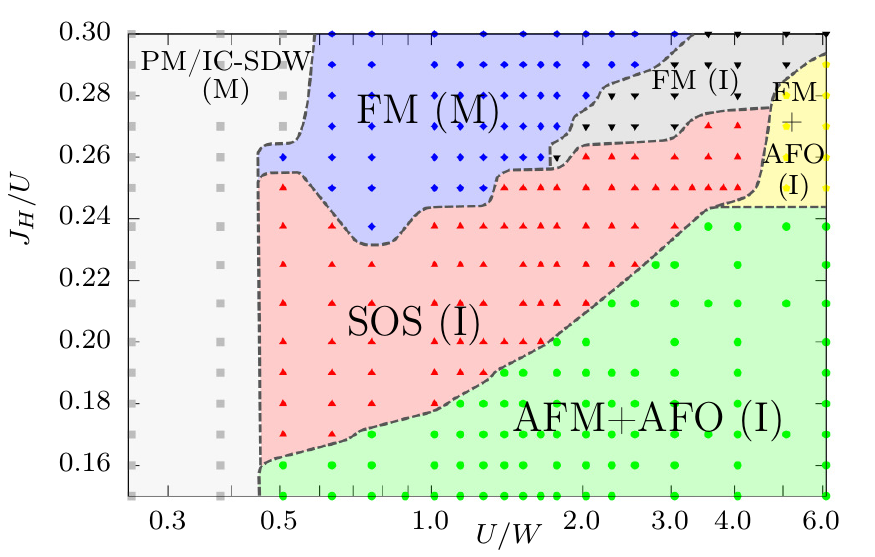}}
\caption{$J_H/U$ vs $U/W$ phase diagram calculated using the Hartree-Fock method for a two-dimensional system. Lattice sizes $12 \times 12$ and $16 \times 16$ were used. Electronic density is $n=2/3$ i.e. two electrons per site. The notation PM, IC-SDW, FM, AFM, AFO, and SOS stands for paramagnetic, incommensurate spin-density wave, ferromagnetic, antiferromagnetic, antiferro-orbital, and staggered orbital stripe order, respectively. I and M stand for insulating and metallic, respectively.
}
\label{Fig3}
\end{figure}

This section discusses the results for two-dimensional clusters
 calculated using the unrestricted real-space Hartree-Fock approximation. 
 The Hartree-Fock decomposition is performed for all the four-fermionic terms in the 
 interaction Eq.~(2) leading to many order parameters $\langle c_{i,\alpha, \sigma}^{\dagger}c_{i,\beta, \sigma{'}} \rangle$ for each site $i$, 
 where $\alpha, \beta$ are orbitals and $\sigma, \sigma{'}$ are spins. 
 We started the iterative process from random initial conditions for the order parameters 
 and self-consistency was reached using the modified Broyden's 
 method~\cite{broyedn}. A chemical potential $\mu$ is tuned to target the required electronic density. 
 To smooth the phase boundaries we also performed Hartree-Fock calculations 
 starting with order parameters corresponding to the ideal representation of the competing phases.
 To identify these phases we have calculated spin-spin correlations, 
 the associated spin structure factor, local spin moments, orbital-resolved local densities, and the overall density of states.

The main result of this section is the $J_H/U$ vs $U/W$ phase diagram,
presented in Fig.~\ref{Fig3}. Calculations were performed for all the points indicated, employing either 
$16 \times 16$ or $12 \times 12$ cluster sizes. 
In the small $U$ region, mainly for $U/W \lesssim 0.5$, as expected we found either a featureless  
paramagnetic metal (PM) or an  incommensurate spin-density-wave metallic (IC-SDW) phase, smoothly 
connected to one another. Because this regime does not seem experimentally relevant, we did not explore this region
in further detail. 

More important for our purposes, in the range $J_H/U < 0.24$ coexisting antiferromagnetism and antiferro-orbital ordering (AFM + AFO) was identified. This state is insulating i.e. it has a
robust gap at the chemical potential in the density of states.  
The value of $U$ needed to stabilize this phase strongly depends on the value of $J_H$ 
(see green region of the phase diagram). 
\begin{figure}[h]
\centering
\rotatebox{0}{\includegraphics*[width=\linewidth]{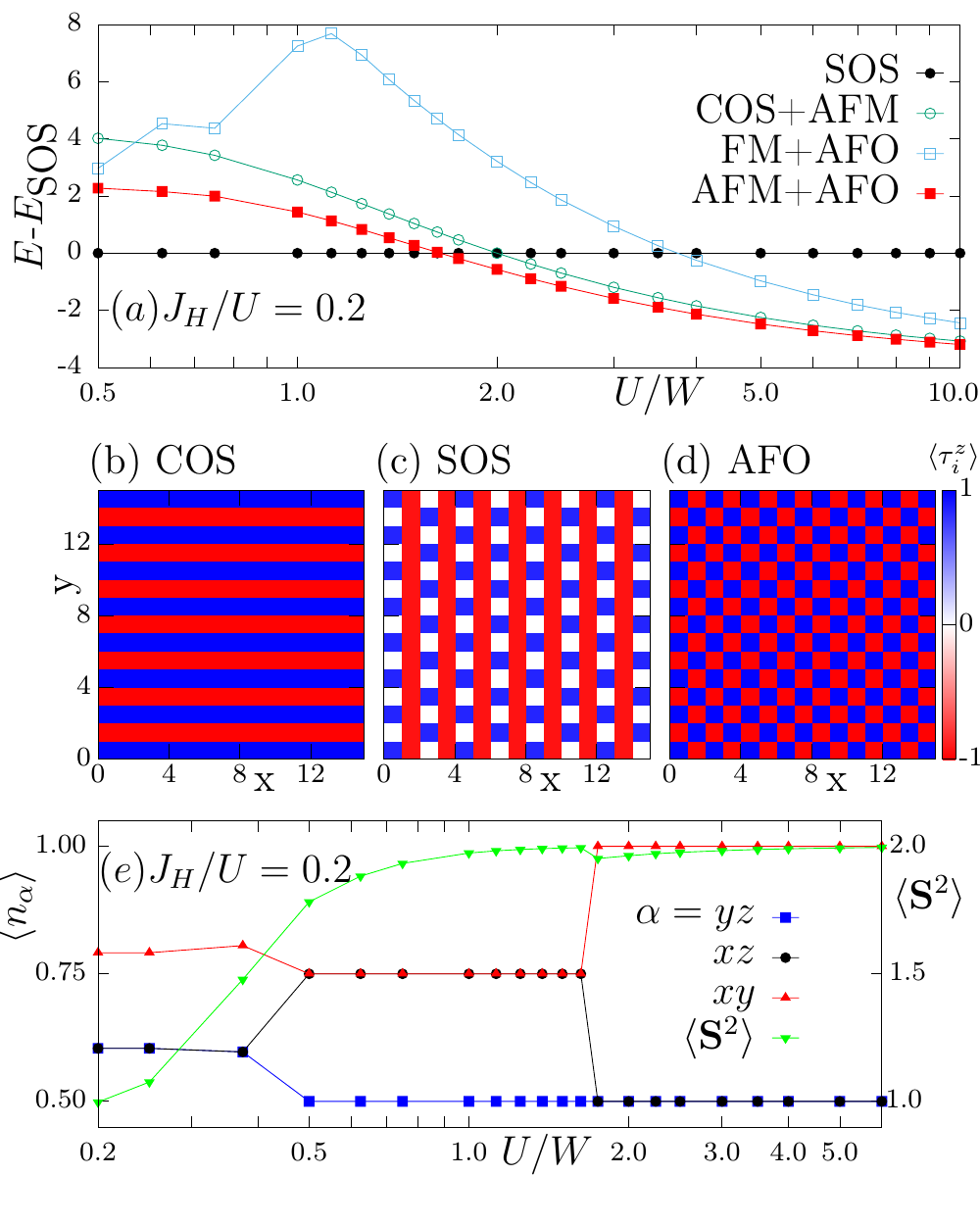}}
\caption{Panel (a) contains the energies of several states (indicated) varying $U/W$, at fixed $J_H/U=0.2$. In panels (b), (c), and (d)
the real-space values of $\langle\tau^{z}_{i}\rangle$ are displayed color coded for the COS, SOS, and AFO states, respectively. See color explanation in text.
In (e), the average orbital-resolved electronic occupation $\langle n_{\alpha} \rangle$ and average local spin moment $\langle \mathbf{S}^{2} \rangle$ is shown for various values of $U/W$.}
\label{Fig4}
\end{figure}

In the large Hund coupling region 
we found ferromagnetic (FM) ordering, driven mainly by double exchange. In this regime several interesting phases
were identified. In particular,
we observed a metal to insulator transition increasing $U$ 
(blue and grey colored regions), and at large $U$ the FM order is accompanied with AFO in a small
portion of the parameter space explored.
From the results in Fig.~\ref{Fig3}, we can safely claim that AFO is stabilized by large $U$~\cite{takashi}.
We have not found (via mean field and in the parameter region studied)
the collinear orbital stripe (COS) state considered to be the ground state in Ref.~\cite{zhu}.
Further work is required to confirm or deny its
existence in the full phase diagram.

 Figure~\ref{Fig4}(a) shows the evolution of energies with $U/W$ 
for various states, with $J_H/U$ fixed to 0.2. In the intermediate $U$ region we found a novel state dubbed
staggered orbital stripe (SOS) as the ground state, which is used as the energy of reference in Fig.~\ref{Fig4}(a).
This novel phase appears in the intermediate $U$ and $J_H$ region
[red region in Fig.~\ref{Fig4}(c)], and in this phase the average occupation in 
orbitals $xy$ and $xz$ is 0.75 each, while $yz$ is 0.50, as shown in Fig.~\ref{Fig4}(e).
On the other hand, in the proposed COS phase and in the AFO phase of our focus here, 
the orbital $xy$ is half-filled (i.e. occupation 1.00) while the orbitals $xz$ and $yz$ are 
quarter filled (occupation 0.50 each). 

The colors in Figs.~\ref{Fig4}(c) and~\ref{Fig4}(d) indicate the mean value of the
local $\tau_{i}^{z}$ for the SOS (depicting stripe order) and AFO 
(depicting staggered order) phases, respectively. For example, in the SOS phase, there are vertical red stripes with local occupations $n_{xz}=n_{xy}=1, n_{yz}=0$, and blue/white staggered stripes with
occupations ($n_{yz}=n_{xy}=1, n_{xz}=0$) and ($n_{xz}=n_{yz}=1, n_{xy}=0$),
respectively. In our calculations, we found that the above state is degenerate 
with the state having horizontal blue stripes and staggered red and white stripes, 
as expected. Increasing the interaction strength, for $U/W \gtrsim 1.8$ 
the AFM (spin staggered) + AFO (orbital staggered) ordering 
becomes the ground state [at very large $U/W$ the
COS+AFM and AFM+AFO states are close in energy, see Fig.~4(a)].
Note the survival only of the red and blue colors, showing that $n_{xy}$ is always 1,
with $n_{xz}=1.0/n_{yz}=0.0$ (red) and $n_{yz}=1.0/n_{xz}=0.0$ (blue) alternating from site to site in a staggered manner.  The AFM+AFO is the most experimentally relevant state
in the phase diagram, and thus our most important result.

Some recent experiments employing pure samples of $\alpha$-Sr$_2$CrO$_4$
suggest that the compound is insulating with antiferromagnetic
spin ordering~\cite{Matsuno:Prl,sakurai} (the type of orbital ordering is still unclear experimentally).
Based on these results, we assume the physical regime for this material 
lies approximately within the $U/W$ and $J_H/U$ pink or green range of the phase diagram. 
Thus, our mean field calculations suggest 
that, from the perspective of magnetism only, the novel SOS phase or the AFM+AFO phase are suitable candidates 
at low temperatures for this material, because the other phases are either FM or PM.

\section{V. DMRG Results}
\begin{figure}[h]
\centering
\rotatebox{0}{\includegraphics*[width=\linewidth]{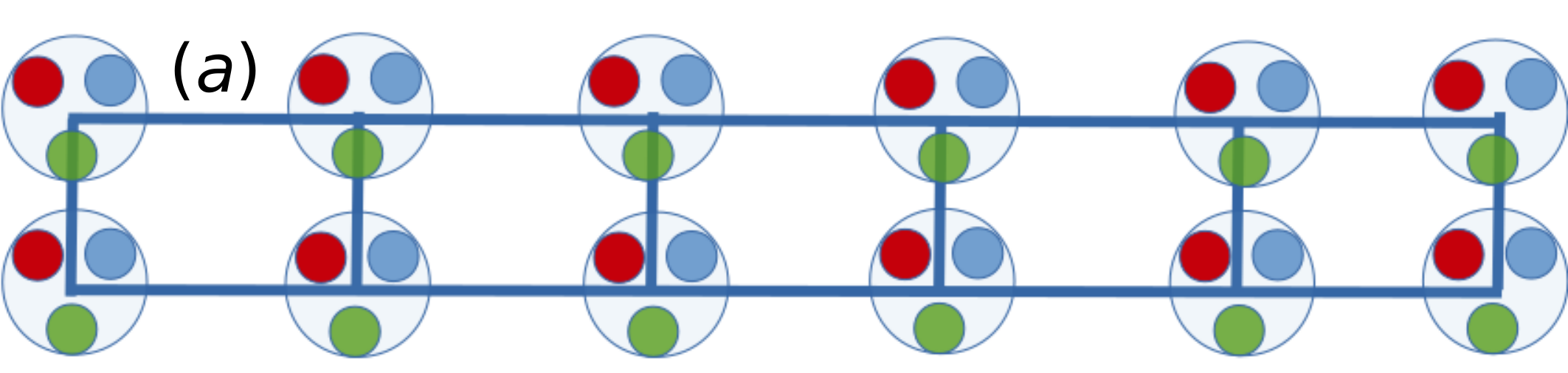}}
\rotatebox{0}{\includegraphics*[width=\linewidth]{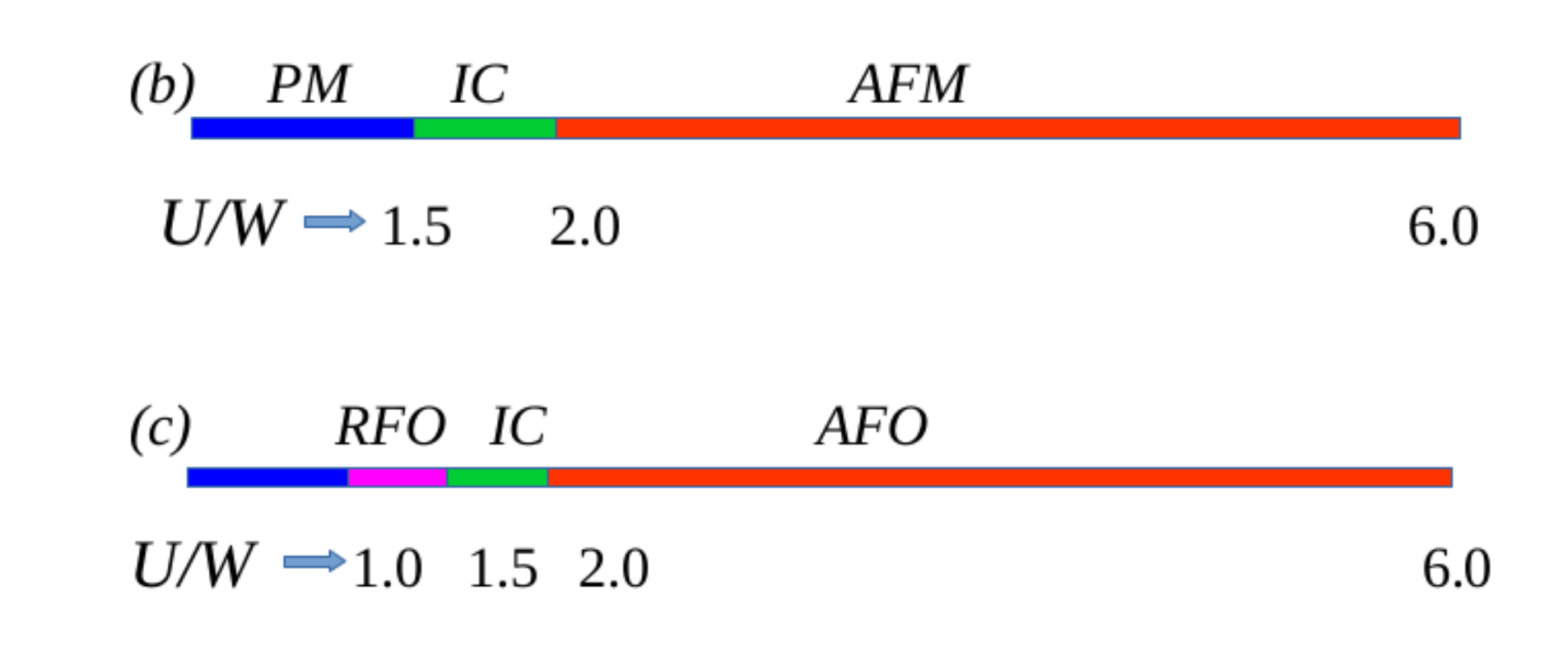}}
	\caption{(a) Schematic representation of a two-leg ladder with three-orbitals at each site.
	 (b,c) DMRG phase diagram at 
	fixed $J_{H}/U=0.2$. (b) contains the spin ordering, where 
	PM stands for paramagnetic phase, IC for incommensurate ordering, and AFM for
	antiferromagnetic staggered spin ordering. (c) contains the orbital ordering, where RFO stands for rung ferromagnetic orbital ordering, and
	AFO for antiferro-orbital ordering along both the leg and rung of the ladder.}
\label{Fig5}
\end{figure}
This section explores the spin and orbital ordering corresponding to a two-leg ladder 
three-orbital Hubbard model [see Fig.~\ref{Fig5}~(a)] employing the density matrix 
renormalization group method. The reason is that several previous examples, such as in models
for Cu- and Fe-based superconductors, has shown that two-leg ladders and planes share qualitatively
many properties~\cite{ladder1,ladder2,ladder3,ladder4,ladder5,ladder6}, 
while DMRG in multiorbital systems can be applied to ladders but not to planes. 
To obtain the physical properties of the proposed model,
we use the previously discussed {\it ab initio} hopping 
parameters of the two-dimensional compound $\alpha$-Sr$_2$CrO$_4$ and vary $U/W$ 
at a fixed $J_H/U=0.2$, because at this $J_H/U$ the Hartree Fock results suggest a rich phase diagram. 

Figures~\ref{Fig5}~(b,c) contain the DMRG magnetic and orbital ordering 
phase diagram for the ground state of the system, based on
DMRG calculations measuring the spin-spin correlation, orbital-resolved site-average
charge density, and orbital-correlation functions. For low values of 
$U/W \lesssim 1.0$, the system is in a paramagnetic phase (PM) without orbital ordering, as expected
in weak coupling. With slight increase in interaction strength $1.0 \lesssim U/W \lesssim 1.5$ 
a rung-ferro-orbital (RFO) type orbital-ordering appears 
(without magnetic ordering). Further increasing the $U$ coupling, at $1.5 \lesssim U/W \lesssim 2.0$ 
incommensurate spin and orbital ordering is observed. While these regions are all interesting from
the fundamental physics perspective, they
will not be the focus of our publication because they are not realized in Sr$_2$CrO$_4$. For this reason, 
these states will not be discussed further. 
\begin{figure}[h]
\centering
\rotatebox{0}{\includegraphics*[width=\linewidth]{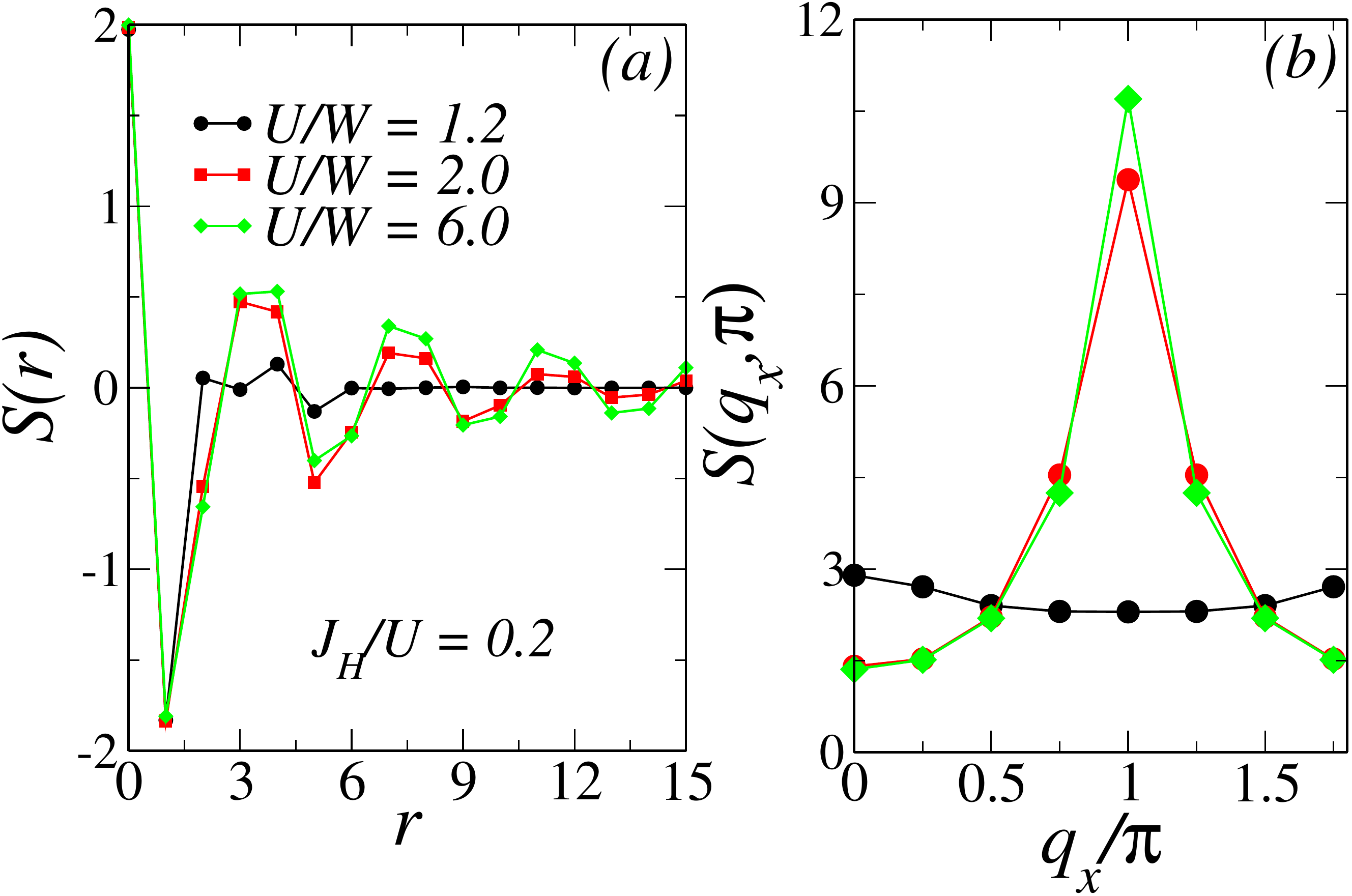}}
\rotatebox{0}{\includegraphics*[width=\linewidth]{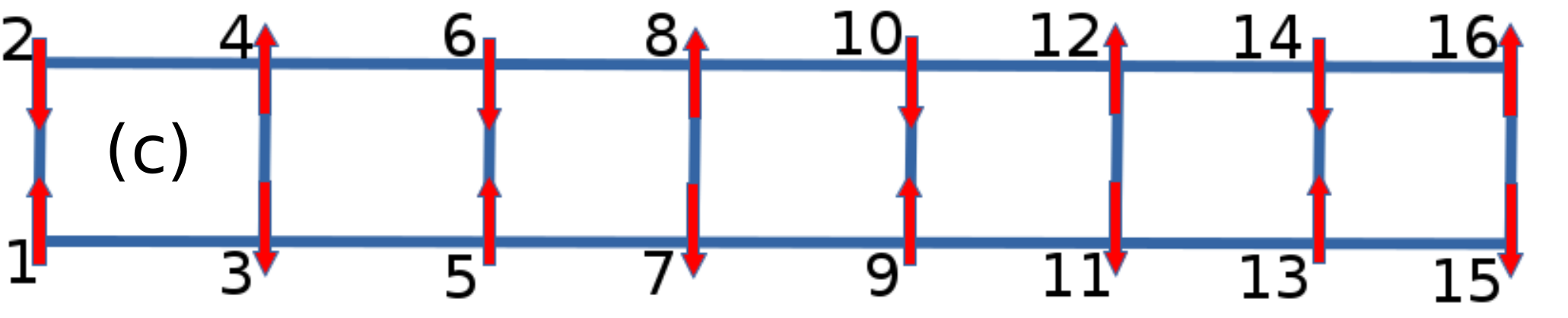}}
	\caption{(a) Real-space spin correlation $S(r)=\langle {\bf{S}}_{1} \cdot 
	{\bf{S}}_j \rangle$ (with $r=|1-j|$), and (b) spin structure factor $S(q_x,\pi)$,
for different values of $U/W$, at a fixed Hund coupling $J_H/U=0.2$, and using a $L=2 \times 8$ cluster. (c)
Schematic of a two-leg ladder, showing the stabilized real-space spin arrangement at large $U$  
with the ``snake'' counting of ladder sites index used in panel (a) to calculate the distance $r$.}
\label{Fig6}
\end{figure}

For $U/W\gtrsim 2.0$, an insulating state with antiferromagnetic (AFM) order
(see sketch in Fig.~\ref{Fig6}) and antiferro-orbital ordering (AFO) (see sketch in Fig.~\ref{Fig8}) becomes stable, results consistent with those of the real-space Hartree-Fock
method. This AFM+AFO phase is our main focus because it is the most experimentally relevant phase for the 
compound $\alpha$-Sr$_2$CrO$_4$.

\subsection{(a) Magnetic Order}

Figure~\ref{Fig6}(a) shows the spin-spin correlation 
$S(r)=\langle {\bf{S}}_1 \cdot {\bf{S}}_j \rangle$ 
vs distance $r$ for different values of $U/W$ and at $J_{H}/U=0.2$. We define 
${\bf{S}}_i=\sum_{\gamma} {\bf{S}}_{i \gamma}$ and in general
$r=|i-j|$, with $i$ and $j$ site indexes, although here we use site $i=1$ as the reference site
to calculate the spin-spin correlation from other sites $j$. 
For $U/W<2.0$, the spin-spin correlation decays very fast with distance $r$,
suggesting a PM phase in the system. 
On the other hand, increasing the on-site repulsion to the range $U/W>2.0$, 
$S(r)$ decays much more slowly, as a power-law, which
is in agreement with having an AFM phase in the system (in one dimension, full long-range order
is not possible). 
As shown schematically in Fig.~\ref{Fig6}, bottom panel,
the system forms antiferromagnetic ordering both along  the legs and along the  
rungs of the ladder. Interestingly, in recent experiments with good quality 
samples of $\alpha-$Sr$_2$CrO$_4$ the presence of AFM ordering 
has been suggested via magnetic susceptibility measurements~\cite{Hsakurai}. 
Also in neutron diffraction studies, a clear AFM staggered order has been unveiled in
the ($a,b$)-plane with a wave vector [1/2,1/2] at low temperatures~\cite{justin}. 

\begin{figure}[h]
\centering
\rotatebox{0}{\includegraphics*[width=\linewidth]{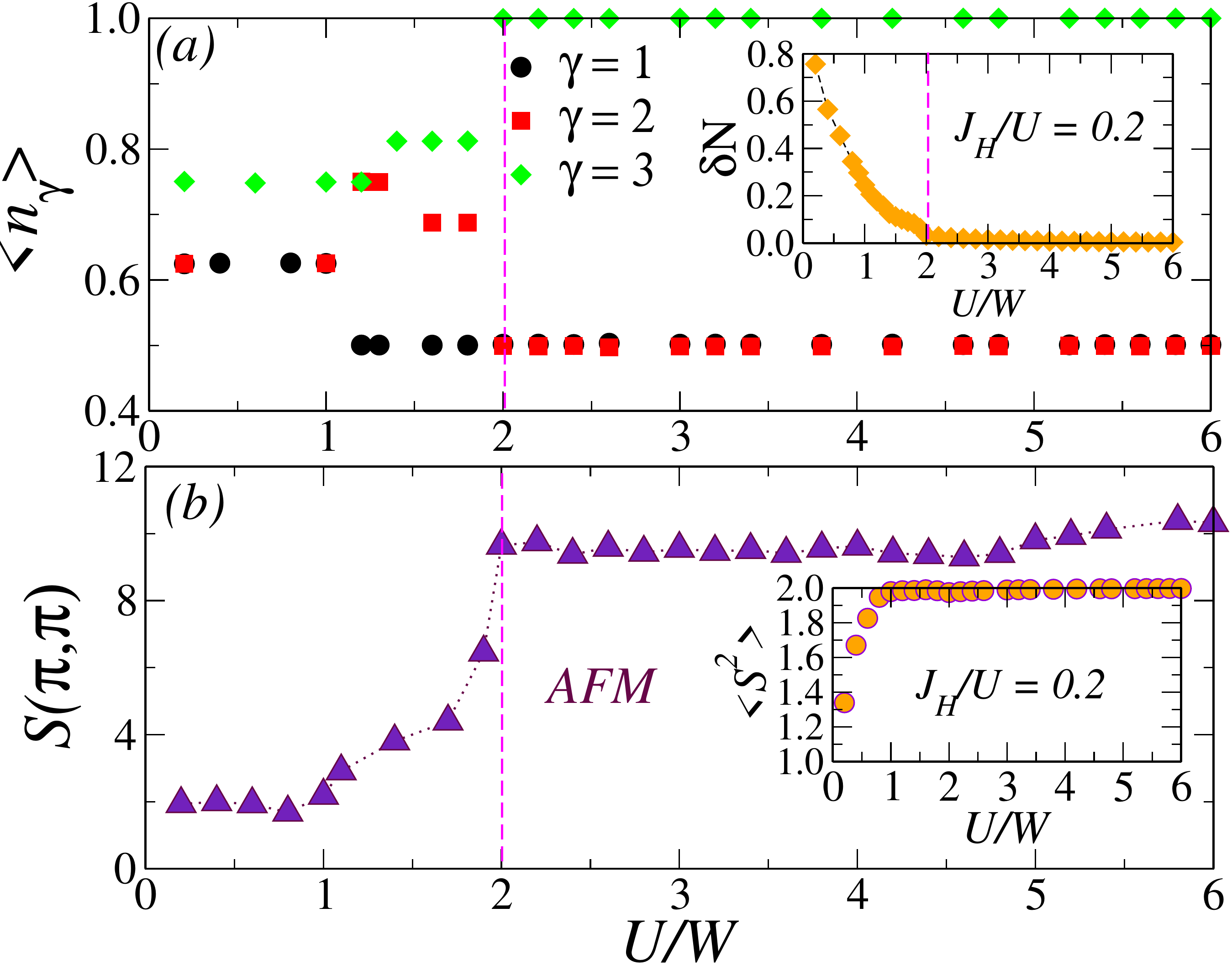}}
\caption{(a) Site-average electron occupancy $n_{\gamma}$ for the three orbitals
	\{$\gamma =1,2,3$\} vs $U/W$. {\it Inset:} site-averaged charge fluctuations $\delta N$
vs $U/W$. (b) Site-average spin structure factor $S(\pi,\pi)$ vs $U/W$. {\it Inset:} 
site-averaged expectation value of the total spin squared vs $U/W$.
These results were obtained using DMRG with cluster size $L=2 \times 8$ at fixed
	$J_H/U=0.2$}
\label{Fig7}
\end{figure}

In order to visualize our results for
magnetic ordering in reciprocal space, we have calculated the spin 
structure factor $S(q_x,q_y)=(1/L) \sum_{i,j} e^{-i \bold{q} \cdot \bold{r}_{ij}} 
\langle {\bf{S}}_i \cdot {\bf{S}}_j\rangle$ 
[in particular, we calculated and used all the correlations $S(r)$ available in our ladder
averaging over all possible distances $r=|i-j|$]. 
Figure~\ref{Fig6}(b) displays the spin structure factor $S(q_x,\pi)$ 
for different values of $U/W$ at fixed $J_H/U=0.2$.
A robust sharp peak at $S(\pi,\pi)$ emerges
for $U/W>2.0$ [as schematically shown in Fig.~\ref{Fig6}(c)]. Interestingly, the sharpness of the peak at $(q_x=\pi,q_y=\pi)$ suggests that 
even using a two-leg ladder, the spin AFM ordering expected in the two-dimensional compound 
Sr$_2$CrO$_4$ can be predicted using DMRG.

Figure~\ref{Fig7}(a) shows the site-average occupancy of 
orbitals $\langle n_{\gamma} \rangle$ vs. $U/W$. 
Interestingly, for $U/W \gtrsim 2.0$
the population of $\gamma=3 $ (the $d_{xy}$ orbital) reaches 1,
while the other two orbitals $\gamma=1$ ($d_{xz}$) and $\gamma=2$ ($d_{yz}$) 
approach $1/2$. The occupancy of orbitals $\langle n_{\gamma} \rangle$
is consistent with results using DFT and real-space Hartree-Fock [see Fig.~\ref{Fig4}(e)].
As discussed in the DFT section, the Jahn-Teller distortion results in 
reverse splitting of the $t_{2g}$ orbitals into lower ($d_{xy}$) and higher degenerate
($d_{xz}$,$d_{yz}$) orbitals. This naturally leads to occupancy 1 for the $d_{xy}$ orbital, 
while the remaining single electron is shared by the degenerate $d_{xz}$ and $d_{yz}$ orbitals. 

In order to find the metallic vs insulating character of the system with 
increasing interaction strength $U/W$, in the inset of Fig.~\ref{Fig7}(a),
 we show the charge fluctuations $\delta N= 1/L \sum_i\left(\langle n^2_i \rangle -\langle n_i \rangle^2 \right)$ 
 vs $U/W$. For $U/W <2$, $\delta N$ has a finite non-zero value, indicating metallic behavior. 
 However,  for $U/W \gtrsim 2.0$, $\delta N$ approaches zero, suggesting insulating behavior for the system.
Figure~\ref{Fig7}(b) presents the peak value of the spin structure factor $S(\pi,\pi)$ 
vs $U/W$. At $U/W \gtrsim 1.0$, $S(\pi,\pi)$ 
starts growing and saturates to a large value  after $U/W \gtrsim 2.0$.
The small values of $S(\pi,\pi)$ and finite $\delta N$, 
indicates with clarity a metallic paramagnetic phase for $U/W \lesssim 1.5$. 
On the other hand, the large values of $S(\pi,\pi)$ for $U/W \gtrsim 2.0$ 
signal a robust insulating state with AFM ordering in the system. 
This evidence of insulating behavior with dominating AFM  $S(\pi,\pi)$ 
ordering  is in excellent agreement with the recent experiments 
based on neutron diffraction measurements for $\alpha$-Sr$_2$CrO$_4$~\cite{sakurai}.

The inset of Fig.~\ref{Fig7}(b) shows the mean value of the local spin-squared 
averaged over all sites  $\langle S^2 \rangle=\frac{1}{L} \sum_i \langle {\bf{S}}_i \cdot {\bf{S}}_i \rangle $. 
For $U/W \gtrsim 1.0$ the local spin moment is fully developed at each site and 
acquires the value $S=1$ (i.e, magnetic moment 2.0 $\mu_B$), primarily driven by a robust Hund coupling.
The  results for the spin structure factor $S(\pi,\pi)$ and $\langle S^2 \rangle$ 
suggest a robust spin $S=1$ antiferromagnetic N\'eel ordering in the system
for $U/W \gtrsim 2.0$.

\subsection{(b) Orbital Order}

As explained in the DFT section, the reverse splitting of $t_{2g}$ orbitals 
into lower ($d_{xy}$) and higher degenerate ($d_{xz}$,$d_{yz}$) 
orbitals opens the possibility of orbital ordering in the system. 
The site-average electronic occupancy of orbitals [$\langle n_{1} 
\rangle = \langle n_{2} \rangle=0.5$ and $\langle n_{3} \rangle = 1 $] 
(see Fig.~\ref{Fig7}(a) for $U/W \gtrsim 2.0$) also hints towards a reverse 
splitting and suggest the presence of an active orbital degree of freedom in the real compound
$\alpha$-Sr$_2$CrO$_4$. 
Using the DFT method, antiferro-orbital ordering has been shown in Ref.~\cite{takashi}
for the compound $\alpha$-Sr$_2$CrO$_4$ [see Appendix Fig.~\ref{Fig12}, where
 using DFT calculations we also obtain antiferro-orbital ordering]. 

\begin{figure}[h]
\centering
\rotatebox{0}{\includegraphics*[width=\linewidth]{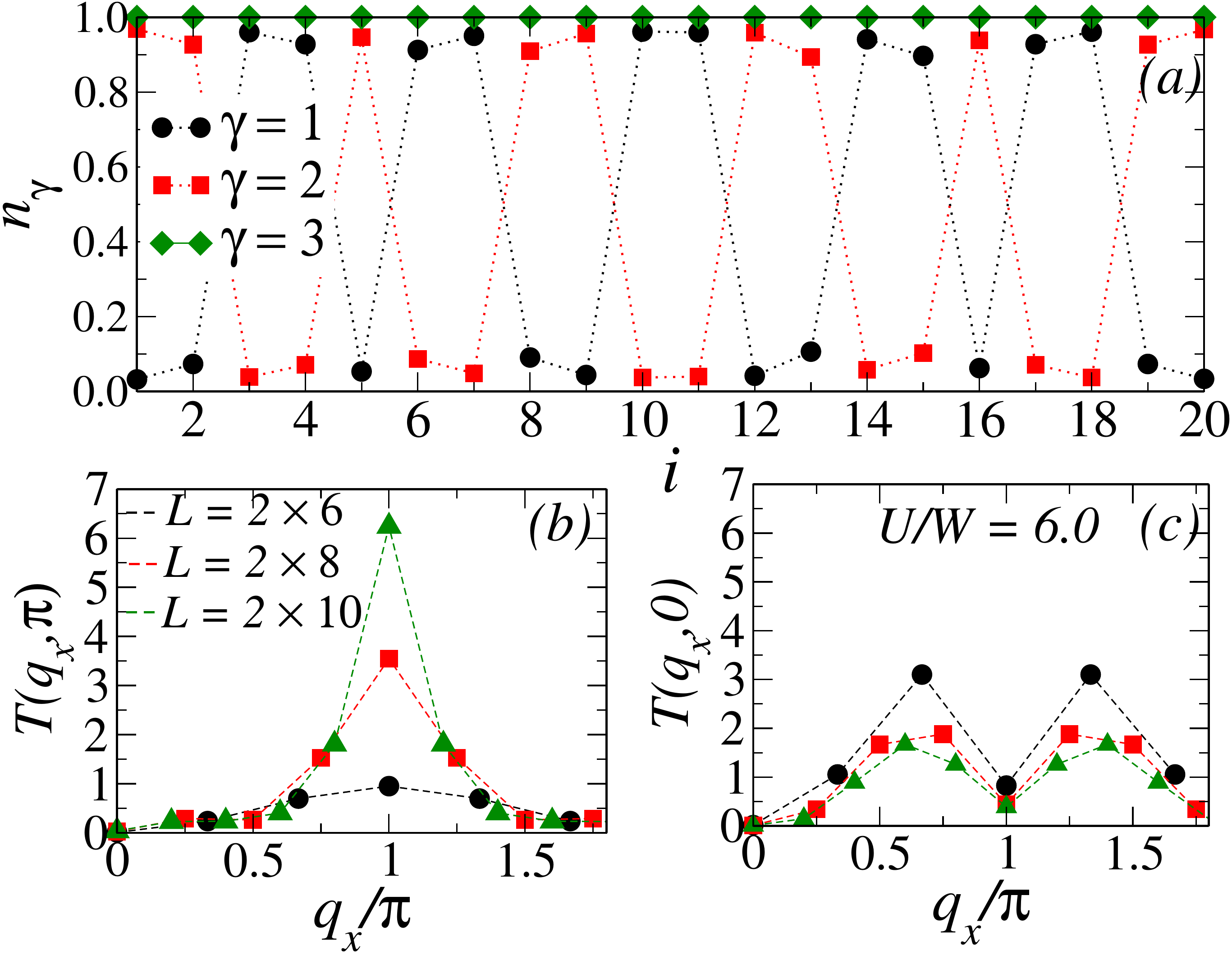}}
\rotatebox{0}{\includegraphics*[width=\linewidth]{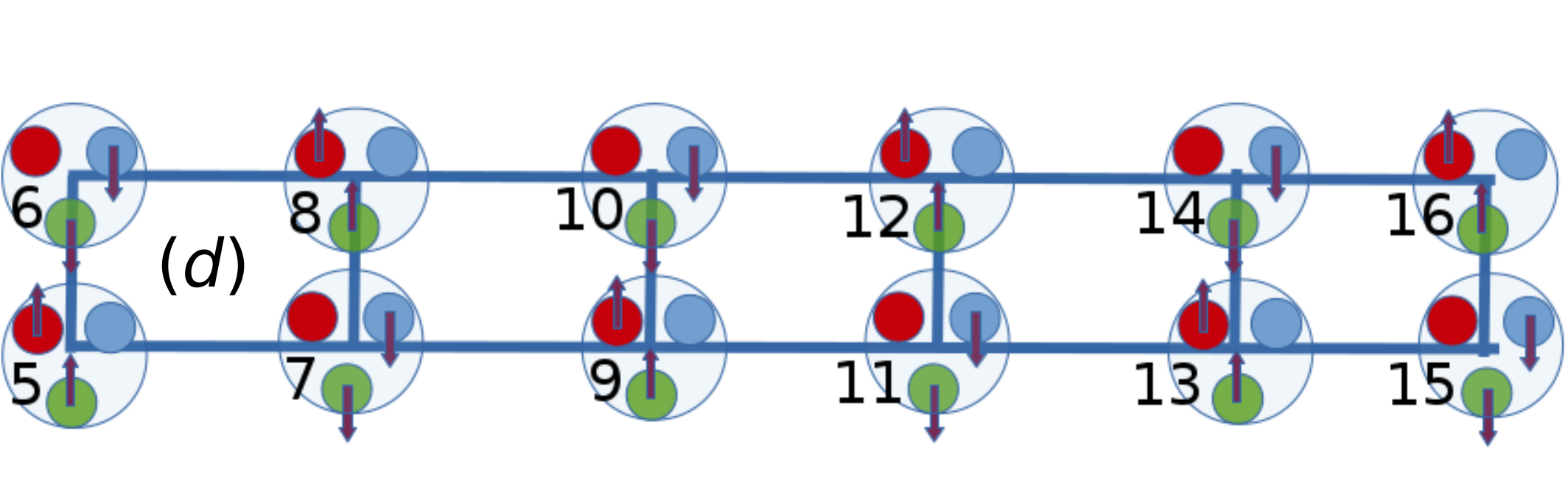}}
	\caption{(a) Electronic charge occupancy $\langle n_{\gamma,i}\rangle$ 
	for the three orbitals $\{ \gamma=1,2,3\}$ vs. site index $i$ at $U/W=6.0$.
	 (b,c) The orbital-ordering structure factors (b) $T(q_x,\pi)$ (c) $T(q_x,0)$
	 at $U/W=6.0$ and $J_H/U=0.2$. In panel (d) is
	the schematic representation of the electronic occupancies for the
	three orbitals $d_{xy}$ (green circles,) $d_{xz}$ (red circles), and $d_{yz}$ (blue circles)
	at each site of a two-leg ladder system.}
\label{Fig8}
\end{figure}

Here, to find the real-space orbital ordering pattern in our ladder model for $\alpha$-Sr$_2$CrO$_4$
we focus on the interaction parameter $U/W \gtrsim 2.0$ (because the model is in 
an insulating state with AFM-spin ordering for $U/W \gtrsim 2.0$).
In Fig.~\ref{Fig8}(a), we plot the population of the three orbitals $ n_{\gamma,i}$ vs.
the site index $i$ at $U/W=6.0$ and $J_H/U=0.2$ for cluster size $L=2 \times 10$.
As shown in Fig.~\ref{Fig8}(a), the orbital 3 ($d_{xy}$) takes value one 
for all sites, whereas orbitals 1 ($d_{xz}$) and 2 ($d_{yz}$) show a dominating 
staggered orbital ordering pattern, both along the rung and legs of the ladder 
[except the first two and last two rungs of the ladder which display ferro-type 
orbital ordering, but likely this is an edge effect due to the open boundary conditions of DMRG]. 
The bottom sketch of Fig.~\ref{Fig8} (panel (d)) illustrates the real-space orbital 
and spin pattern on the two leg ladder from a 2$\times$10 cluster. Note that orbital $d_{xy}$ (green circle) is
always singly occupied, while orbitals $d_{xz}$ (red circle) and  $d_{yz}$ 
(blue circle) are occupied on alternate sites along the 
rungs and legs of the ladder (namely AFO-orbital ordering is shown). 

To gather more insight, we investigate the system size dependent 
orbital-ordering structure factor 
$T(q_x,q_y)=(1/L)\sum_{i,j} e^{-i {\bf{q}} \cdot {\bf{r}}_{ij}} \langle T_i  T_j\rangle$, [where $T_i=n_{\gamma=1,i}-n_{\gamma=2,i}$] at  $U/W=6.0$ for two wavevectors $q_y=0$ and $\pi$.
In Fig.~\ref{Fig8}(b) and Fig.~\ref{Fig8}(c), we show the orbital-ordering structure 
factors, $T(q_x,\pi)$ and $T(q_x,0)$ vs. $q_x$, respectively, for three cluster sizes $L= 2 \times 6$ , $2 \times 8$ and $2 \times 10$, at $U/W=6.0$ and $J_H/U=0.2$.
Interestingly, we find the value of the peak at $q_x=\pi$ for $T(q_x,\pi)$ increases 
sharply with increasing the system size, see Fig.~\ref{Fig8}(b).
However, the peak values for $T(q_x,0)$ 
decreases with increasing the system size. Thus, the latter peak is probably 
due to finite size effects and it can be discarded.
The increase in strength of the ($q_x=\pi$, $q_y=\pi$) peak clearly indicates that
for large system sizes the antiferro-type orbital ordering will be dominating 
for $U/W \gtrsim 2.0$.

\subsection{(c) Lanczos results and density of states}
\begin{figure}[h]
\centering
\rotatebox{0}{\includegraphics*[width=\linewidth]{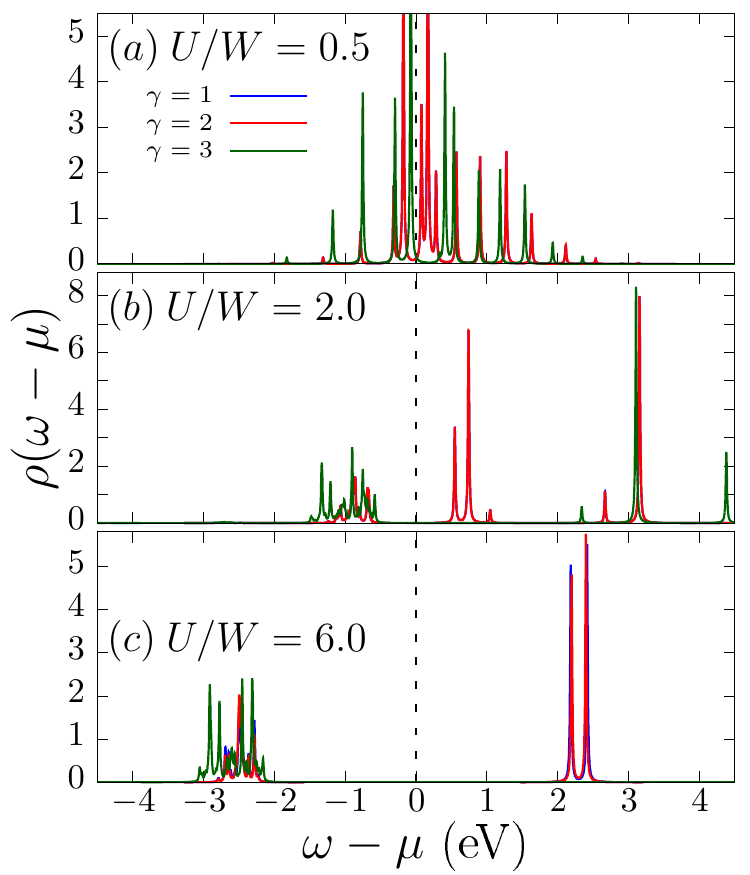}}
	\caption{ Orbital-resolved density of state (DOS) vs. $\omega - \mu$ for different values
	of interaction strengths (a) $U/W=0.5$, (b) $U/W=2.0$, and (c) $U/W =6.0$,
	at fixed $J_H/U=0.2$, using the Lanczos diagonalization method for a small
	$L=2 \times 2$ cluster.}
\label{Fig9}
\end{figure}


To characterize the metallic vs insulating behavior of the system 
varying the interaction strength, in addition to the charge fluctuations previously discussed 
we have also investigated the orbital-resolved density of state using the Lanczos method
for a small cluster $ L =2 \times 2$. Although the cluster is small, the results are enough to
explain qualitatively the metallic and insulating behavior of the system.
Figure~\ref{Fig9} contains the orbital-resolved density of states (DOS)
vs $\omega - \mu$ ($\omega$ is the frequency and $\mu$ is the chemical potential) 
for three values of $U/W =0.5$, $2.0$, and $6.0$, and at fixed $J_H/U=0.2$.
As shown in Fig.~\ref{Fig9}(a), all the three orbitals carry 
non-zero weight at $U/W=0.5$, indicating metallic behavior. 
However, in Figs.~\ref{Fig9}(b) and~\ref{Fig9}(c) the system opens a large gap,
compatible with insulating behavior at $U/W =2$ and $6$. 
The insulating behavior of thin films of the compound Sr$_2$CrO$_4$ has been experimentally
demonstrated by measuring the optical conductivity spectra~\cite{Matsuno:Prl}.

\section{VI. Origin of staggered AFM and AFO order}
\begin{figure}[h]
\centering
\rotatebox{0}{\includegraphics*[width=\linewidth]{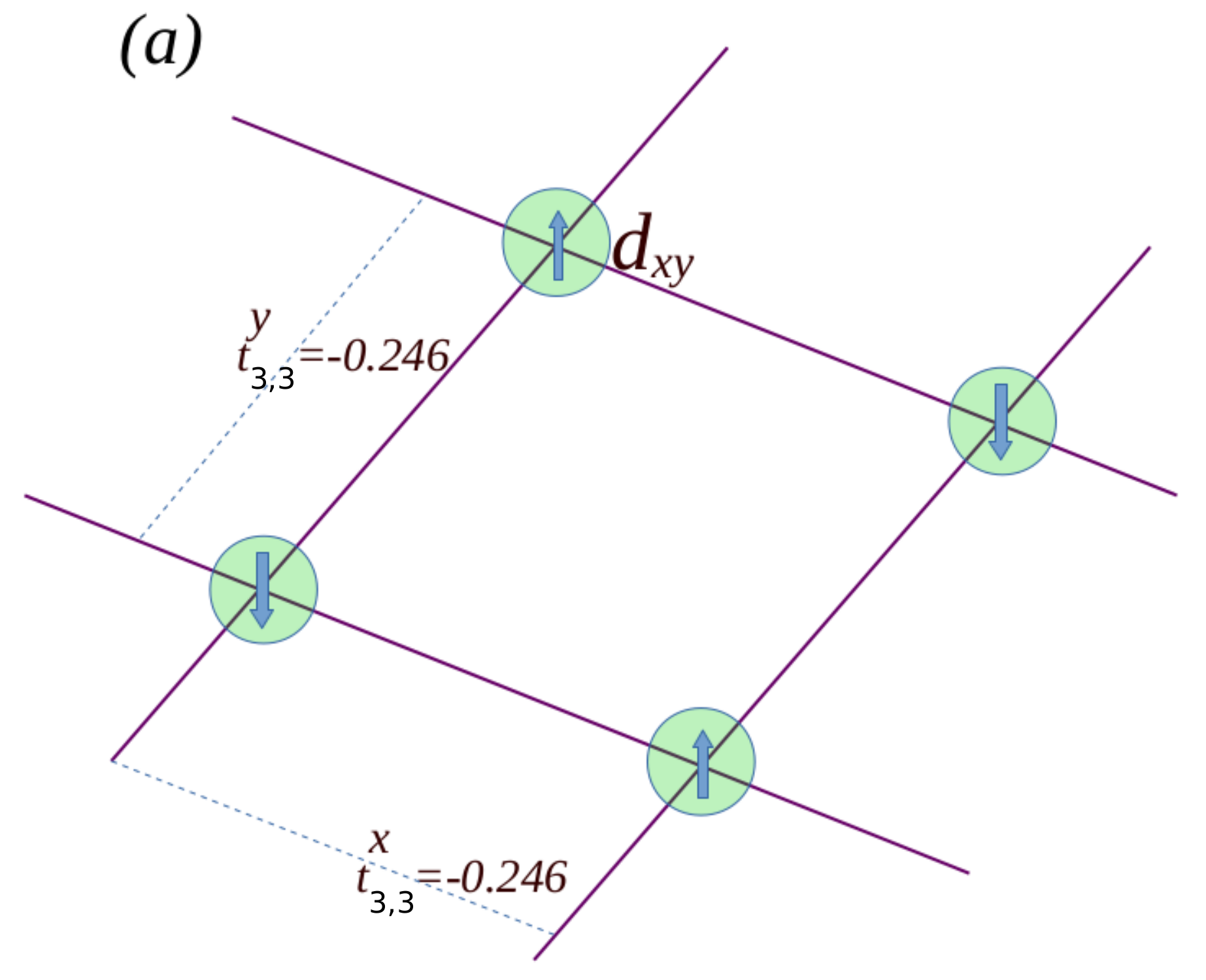}}
\rotatebox{0}{\includegraphics*[width=\linewidth]{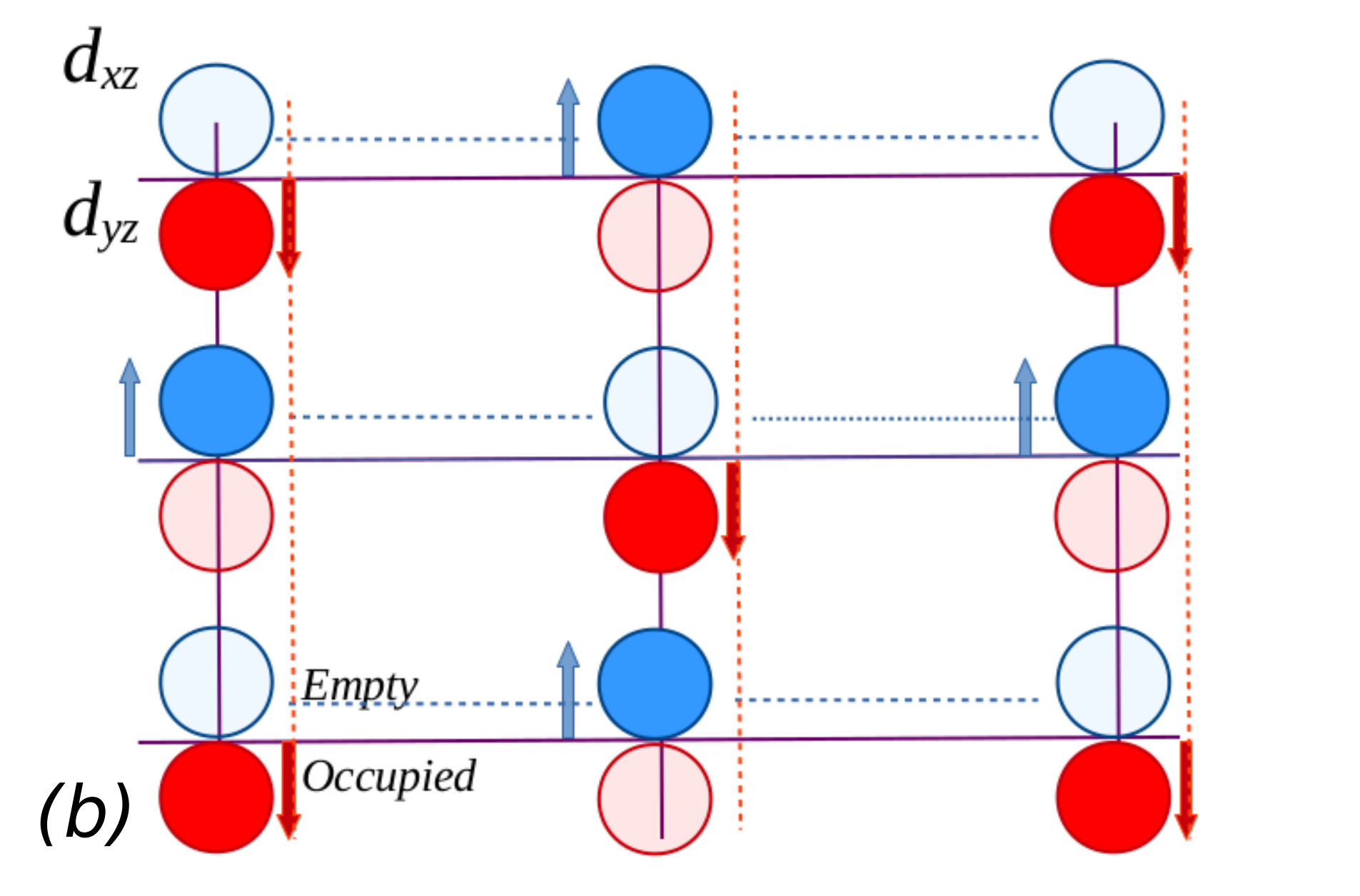}}
	\caption{(a) Sketch representing the $d_{xy}$ orbitals, which have the largest hopping amplitude 
	along the $x$ and $y$ directions in the $t_{2g}$ sector. This orbital is always occupied by one electron, thus it develops staggered spin ordering.
	(b) Schematic representation of the $d_{xz}$ (blue) and $d_{yz}$ (red) orbitals. Circles in dark color 
represent occupied orbitals while light color are empty orbitals. The hopping amplitudes for $d_{xz}$ orbitals along the $x$ direction ($t^x_{1,1}=-0.193$) dominate over the
	$y$ direction ($t^y_{1,1}=-0.039$). Reciprocally, the hopping amplitudes for the $d_{yz}$ orbitals along $y$ 
($t^y_{2,2}=-0.193$) dominate over the $x$ direction hopping ($t^x_{2,2}=-0.039$). Those dominant hoppings are represented by colored dashed lines.}
\label{Fig10}
\end{figure}

As discussed in previous sections, our numerical results (HF, DMRG, and DFT) 
predict an insulating antiferromagnetic state with antiferro-orbital ordering for the three-orbital
model representing the compound $\alpha$-Sr$_2$CrO$_4$.  
We here provide an intuitive explanation for the existence of this spin and orbital arrangement.

The stability of AFM order at $U/W \gtrsim 2.0$ can be explained intuitively by considering the dominant
role of the most mobile orbital $d_{xy}$ together with the on-site interaction $U$. The $d_{xy}$ orbital
is separated from the rest of the orbitals by the crystal field, and it has the largest hopping amplitude
along both the $x$ and $y$ directions, see sketch in Fig.~\ref{Fig10}(a). Thus, as a first crude approximation
we can focus on this orbital. Its half-filled nature, one electron per $d_{xy}$ orbital,
makes this subspace effectively a one-orbital Hubbard model at half-filling $n=1$.
Because of the large on-site interaction $U$, which generates an effective Heisenberg superexchange model, 
staggered AFM order dominates. Moreover, because of the robust on-site Hund interaction $J_{H}$,
 the electrons located in the other $d_{xz}$ and $d_{yz}$ orbitals will follow the 
same spin pattern as the $d_{xy}$ orbital. Thus, the driver of the AFM order is the $d_{xy}$ orbital.

\begin{figure}
\centering
\includegraphics[width=0.48\textwidth]{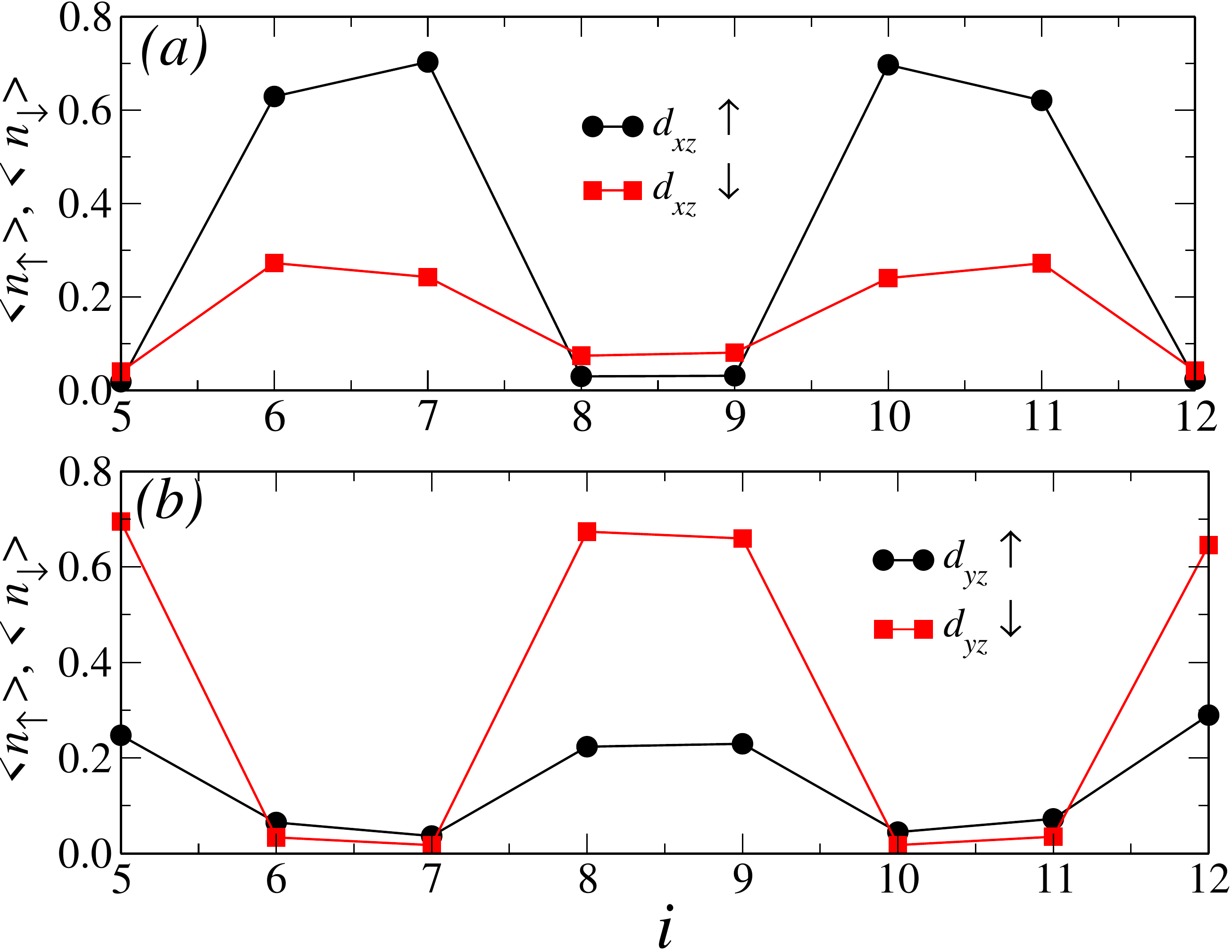}
	\caption{Spin-resolved charge occupancy of the $d_{xz}$ and $d_{yz}$ orbitals vs site index $i$, using DMRG applied to a
	two-leg ladder of size $L=2 \times 8$. Panel (a) shows that $\langle n_{i,1, \uparrow}\rangle >  \langle n_{i,1, \downarrow}\rangle $ 
	for the $d_{xz}$ orbitals. Panel (b) shows that $\langle n_{i,2, \uparrow}\rangle <  \langle n_{i,2, \downarrow}\rangle $ 
	for the $d_{yz}$ orbitals, in agreement with the qualitative description presented in Sec.~VI.}
\label{Fig11}
\end{figure}

The existence of antiferro-orbital ordering in the twofolded degenerate space of orbitals $d_{xz}$
and $d_{yz}$ can be explained by considering their hopping amplitudes
(different along the $x$ and $y$ directions) and the on-site interorbital repulsion $U'=U-2J_{H}$.
Note that the hopping amplitude for $d_{xz}$ orbitals is much larger
along the $x$ direction than $y$, while for $d_{yz}$ the reciprocal occurs,
i.e. much larger along $y$ than $x$. To minimize the $U'$ repulsion it is natural to spread the
charge in this $d_{xz}-d_{yz}$ sector, leading to one electron per site in this subspace. 

Let us arrange electrons in the $d_{xz}-d_{yz}$ subspace starting at the center site in the 3$\times$3 lattice shown in Fig.~\ref{Fig10}(b).
Arbitrarily, let us place there an electron with spin down in orbital $d_{yz}$, indicated by a filled red circle. Because this orbital has hopping primarily along the $y$ axis, then to help with the electronic itineracy, 
which reduces the energy via the tight-binding term, the two neighboring sites along $y$ should not have electrons
in $d_{yz}$. Then, in those sites the electron is located in the $d_{xz}$ orbital, indicated with a filled blue
circle. Because the spin must follow the pattern dictated by the $d_{xy}$ orbital due to J$_H$, then those electrons have spin up.

Consider now the upper row in Fig.~\ref{Fig10}(b). After the central spin is fixed in orientation and orbital 
$d_{xz}$ location by the discussion above,
by the same rationale as in the previous paragraph, then the two sites left and right must contain an electron
in the $d_{yz}$ orbital with spin pointing down. By this procedure all the sites of the lattice can be filled, and
the pattern that emerges is in Fig.~\ref{Fig10}(b). Clearly, the orbital $d_{xz}$ is polarized up and has a staggered
occupancy in the two dimensional lattice, and the orbital $d_{yz}$ is the opposite, namely polarized down occupying 
the other half of the lattice. Electrons in $d_{xz}$ move primarily along the $x$ direction, while those in $d_{yz}$
move along the $y$ direction. Thus, both of them are effectively one dimensional with regards to their mobility, while the $d_{xy}$ electrons are two dimensional. With this arrangement, the dominant $U$ repulsion is not active since
there is no double occupancy, the $U'$ repulsion is minimized by spreading the charge, the $J_H$ term which is
effectively attractive is active in all sites, and the kinetic energy is optimized because all electrons in
the $d_{xz}$ and $d_{yz}$ subspaces can jump to empty nearest-neighbor sites.


The spin polarization of the $d_{xz}$ and $d_{yz}$ orbitals emerging from this description was 
further confirmed by DMRG calculations on the two-leg ladder. We find that the electronic density
for the $d_{xz}$ orbitals satisfies $\langle n_{i,1,\uparrow} \rangle > \langle n_{i,1,\downarrow}\rangle$, 
 while for the $d_{yz}$ orbital $\langle n_{i,2,\uparrow}\rangle < \langle n_{i,2,\downarrow}\rangle$
 (see Fig.~\ref{Fig11}). This pattern of spins up in $d_{xz}$ and spin down in $d_{yz}$
can be reversed, producing a degeneracy two in the state.

\section{VII. Conclusions}

In this publication, the magnetic and orbital ordering of the compound 
$\alpha$-Sr$_2$CrO$_4$ has been investigated by using unrestricted 
real-space Hartree-Fock, DMRG, and Lanczos techniques. 
Realistic hopping amplitudes for the three-orbital Hubbard model used here
were derived using {\it ab initio} calculations. 
We applied the Hartree-Fock method to the two-dimensional three-orbital Hubbard model,
and we found a rich phase diagram, with a variety of ferromagnetic (FM), antiferromagnetic (AFM),
staggered orbital (SOS), and antiferro-orbital (AFO) ordered phases.
Furthermore, using DMRG for two-leg ladders we also investigated the spin and orbital
ordering with the same realistic hopping parameters corresponding to $\alpha$-Sr$_2$CrO$_4$.
Both the Hartree-Fock and DMRG methods predict the same insulating ground state 
with antiferromagnetic spin ordering, in excellent agreement with experiments.
Moreover, the unique reverse splitting of $t_{2g}$ orbitals
 for the compound Sr$_2$CrO$_4$ unveiled by DFT is important to understand the orbital ordering. Both our
 Hartree-Fock and DMRG results converge to a stable antiferro-orbital ordering for
 moderate to large interaction strength $U$, a range expected to be relevant for
 the real material Sr$_2$CrO$_4$. Using the Lanczos method for a small
 size cluster, the orbital-resolved density of state was calculated, and it displays
 insulating behavior for this system.  We believe that 
 our numerical results related to spin and orbital ordering,
 using a realistic three-orbital Hubbard model, provide a qualitatively accurate description
 for the compound Sr$_2$CrO$_4$. With the evidence provided here and in other related publications 
that the orbital degree of freedom is active in Cr oxides, a plethora of attractive possibilities open up,
such as replicating with Cr the wide variety of orbitally ordered states reported in manganites~\cite{orbmanga1,orbmanga2} 
and ruthenates~\cite{orbruthe}, the effect of strain~\cite{strainmanga,cengiz}, and the possibility of
block states~\cite{block1,block2,block3,block4} or even spirals~\cite{spiral}.
Recent theoretical work has even suggested that superconductivity is possible upon doping a doubly degenerate multiorbital system in chains~\cite{patelnpj} and, thus, similar results in planes could occur.

\section{Acknowledgments}
The work of B.P, Y.Z, N.K, L.-F.L., and E.D. was supported by the U.S. Department of
Energy (DOE), Office of Science, Basic Energy Sciences
(BES), Materials Sciences and Engineering Division. 
G.A. was partially supported by the Center for Nanophase Materials Sciences, 
which is a U.S. DOE Office of Science User Facility, and by the Scientific Discovery  through  Advanced  Computing  (SciDAC) program  funded  by  U.S.  DOE,  Office  of  Science, Advanced  Scientific  Computing  Research  and  Basic Energy Sciences, Division of Materials Sciences and Engineering.
Validation and some computer runs were conducted at the Center for Nanophase Materials Sciences, which is a DOE Office of Science User Facility.

\section{Appendix}
\begin{figure}
\centering
\includegraphics[width=0.48\textwidth]{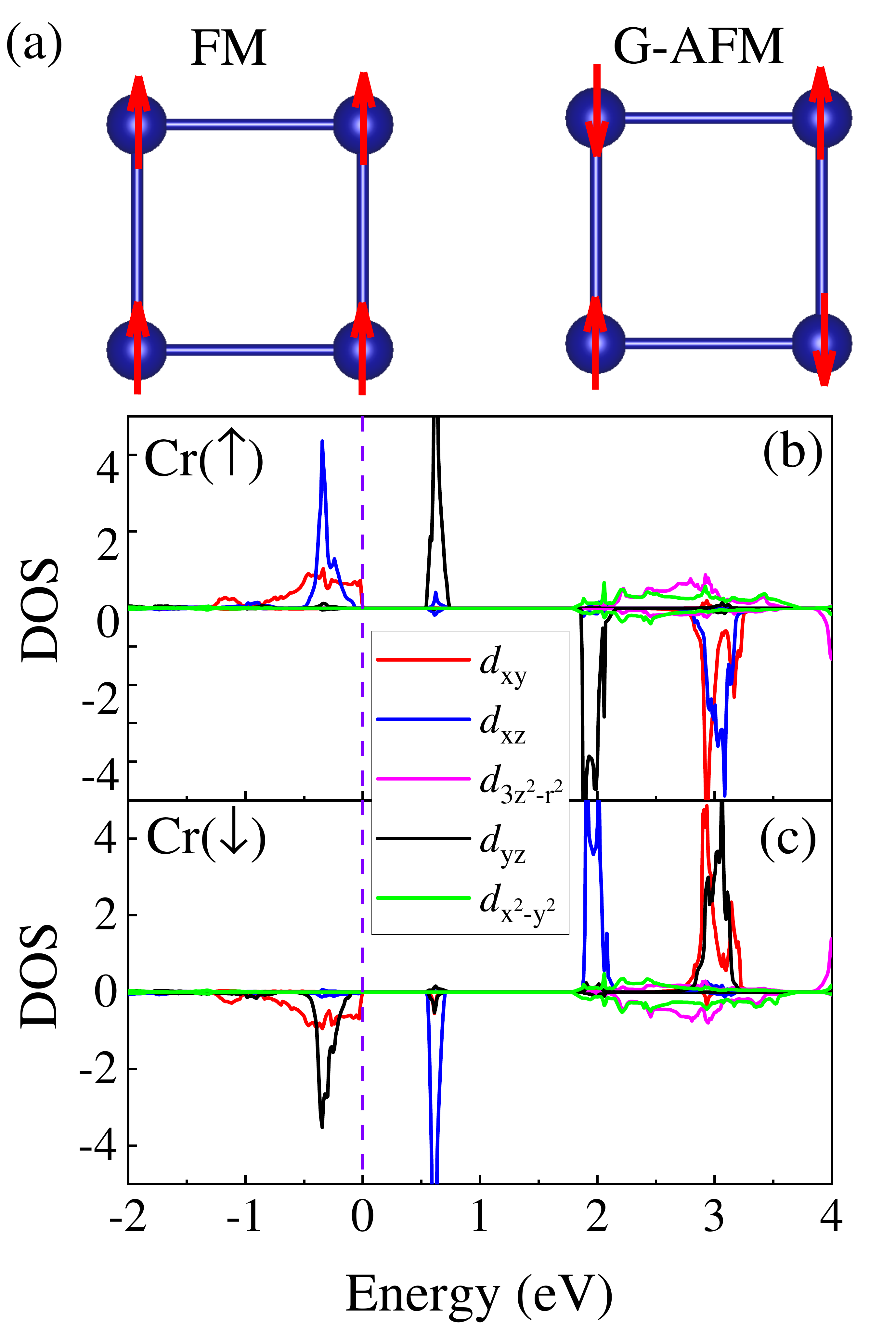}
\caption{(a) Sketch of FM and G-AFM spin configurations in the 2D square lattice, considered in the DFT calculations. Spin-up and spin-down are indicated by arrows. (b-c) Cr-projected local DOS corresponding to the spin-up and spin-down Cr atoms in one plane with a G-AFM type magnetic configuration, respectively. The Fermi level is indicated with dashed lines.}
\label{Fig12}
\end{figure}


In this Appendix, we will discuss the magnetic ground state and orbital ordering state of 
$\alpha$-Sr$_2$CrO$_4$ based on DFT calculations. Considering the $C_4$ symmetry 
of the $2D$ square lattice of Sr$_2$CrO$_4$, we only calculated two possible spin configurations [see Fig.~\ref{Fig12}(a)] by using LSDA+$U_{\rm eff}$, with $U_{\rm eff}$ = 2~eV. We found that the G-AFM ($\pi$, $\pi$) has lower energy than the FM state, 
which is consistent with the experimental results~\cite{zhu,justin} 
and also with our DMRG and real-space Hartree-Fock results in a robust region of parameter space.
As shown in Figs.~\ref{Fig12}(b-c), the orbital ordering physics was also successfully 
realized, namely the $3d_{xy}$ orbitals are occupied in both the spin-up and spin-down channels of the 
Cr atoms, while the $3d_{xz}$/$3d_{yz}$ are occupied in spin-up 
or spin-down Cr atoms, respectively. In this case, the orbital-ordered state 
should have wavevector ($\pi$, $\pi$), namely antiferro orbital ordering (AFO) 
along both $x$ and $y$ directions. Based on our DFT calculations, 
we qualitatively obtained the orbital ordering driven by electronic 
correlation, in excellent agreement with the results more systematically discussed in this publication
using the Hartee-Fock and DMRG calculations. 
Furthermore, we also found a Mott gap in the DOS, as displayed in 
Figs.~\ref{Fig12}(b-c), indicating that the system is a Mott-Hubbard-type insulator in agreement with 
experiments~\cite{Matsuno:Prl}.

In summary, using the DFT calculations, we properly reproduced the recent developments in the experimental study of the Sr$_2$CrO$_4$ system. We showed that the crystal field and Jahn-Teller distortion lead to the ($d_{xy}$)$^1$($d_{xz}$,$d_{yz}$)$^1$ electron occupation, corresponding to a CrO$_6$ octahedral with a 3$d^2$ configuration. Furthermore, we obtained the G-AFM ($\pi$, $\pi$) ground state and ($\pi$, $\pi$) antiferro orbital ordering.


\begin{thebibliography}{99}

\bibitem{kugel1} K. I. Kugel and D. I. Khomskii, {Zh. Eksp. Teor. Fiz. {\bf{64}},
1429 (1973) [Sov. Phys. JETP {\bf{37}}, 725 (1973)]}
\url{http://www.jetp.ac.ru/cgi-bin/e/index/e/37/4/p725?a=list}

\bibitem{kugel} K. I. Kugel and D. I. Khomskii,  {Sov. Phys. Usp. {\bf{25}} 231 (1982).}
\url{https://doi.org/10.1070/PU1982v025n04ABEH004537}

\bibitem{dagotto2001} E. Dagotto, T. Hotta, and A. Moreo, {Phys. Rep. {\bf{344}}, 1 (2001).}
\url{https://www.sciencedirect.com/science/article/abs/pii/S0370157300001216?via%3Dihub}

\bibitem{johnston2010} D. Johnston, Adv. Phys. {\bf{59}}, 803–1061 (2010).
\url{https://doi.org/10.1080/00018732.2010.513480}

\bibitem{pdai} P. Dai, J. Hu, and E. Dagotto, {Nat. Phys. {\bf{8}}, 709 (2012).}
\url{https://doi.org/10.1038/nphys2438}

\bibitem{elbio} E. Dagotto, {Rev. Mod. Phys. {\bf{85}}, 849 (2013).}
\url{https://doi.org/10.1103/RevModPhys.85.849}

\bibitem{chubukov} R. M. Fernandes and A. V. Chubukov, {Rep. Prog. Phys. {\bf{80}}, 014503 (2017).} 
\url{https://iopscience.iop.org/article/10.1088/1361-6633/80/1/014503}
 
\bibitem{scalapino} D. J. Scalapino, Rev. Mod. Phys. {\bf{84}}, 1383 (2012).
\url{https://doi.org/10.1103/RevModPhys.84.1383}

\bibitem{dagotto94} E. Dagotto, {Rev. Mod. Phys. {\bf{66}}, 763 (1994).}
\url{https://doi.org/10.1103/RevModPhys.66.763}


\bibitem{Maeno}Y. Maeno, H. Hashimoto, K. Yoshida, S. Nishizaki, T. Fujita, J. G. Bednorz, and F. Lichtenberg, {Nature  {\bf{372}}, 532–534 (1994).}
	\url{https://doi.org/10.1038/372532a0}


\bibitem{San} L. Ortega-San-Martin, A. J. Williams, J. Rodgers, J. P. Attfield, G. Heymann, and H. Huppertz, {Phys. Rev. Lett. {\bf{99}}, 255701 (2007).}
\url{https://doi.org/10.1103/PhysRevLett.99.255701}

\bibitem{ezhov} S. Yu. Ezhov, V. I. Anisimov, D. I. Khomskii, and G. A. Sawatzky, {Phys. Rev. Lett. {\bf{83}}, 4136 (1999).}
\url{https://doi.org/10.1103/PhysRevLett.83.4136}

\bibitem{tokura}Y. Tokura, N. Nagaosa, {Science {\bf{288}}, 462-468 (2000).}
\url{https://doi.org/10.1126/science.288.5465.462}




\bibitem{komarek} A. C. Komarek, S. V. Streltsov, M. Isobe, T. M\"{o}ller, M. Hoelzel, A. Senyshyn, D. Trots, M. T. Fern\'{a}ndez-D\'{i}az, 
T. Hansen, H. Gotou, T. Yagi, Y. Ueda, V. I. Anisimov, M. Gr\"{u}ninger, D. I. Khomskii, and M. Braden {Phys. Rev. Lett. {\bf{101}}, 167204  (2008).}
\url{https://doi.org/10.1103/PhysRevLett.101.167204}

\bibitem{Jun}  J. Akimitsu, H. Ichikawa, N. Eguchi, T. Miyano, M. Nishi, and K. Kakurai, {J. Phys. Soc. Jpn. {\bf{70}}, 3475 (2001).}
\url{https://doi.org/10.1143/JPSJ.70.3475}

\bibitem{Varignon} N. C. Bristowe, J. Varignon, D. Fontaine, E. Bousquet, and Ph. Ghosez, {Nat. Commun. {\bf{6}}, 6677 (2015).}
\url{https://doi.org/10.1038/ncomms7677}

\bibitem{khoms} D. I. Khomskii and G. A. Sawatzky, {Solid State Commun. {\bf{102}}, 87 (1997).}
\url{https://doi.org/10.1016/S0038-1098(96)00717-X}

\bibitem{Jean} J. Jeanneau, P. Toulemonde, G. Remenyi, A. Sulpice, C. Colin, V. Nassif, E. Suard, E. S. Colera, G. R. Castro, F. Gay, C. Urdaniz, R. Weht, C. Fevrier, 
A. Ralko, C. Lacroix, A. A. Aligia, and M. N. Regueiro, {Phys. Rev. Lett. {\bf{118}}, 207207  (2017).}
\url{https://doi.org/10.1103/PhysRevLett.118.207207}

\bibitem{helman} A. A. Aligia and C. Helman, {Phys. Rev. B {\bf{99}}, 195150 (2019).}
\url{https://doi.org/10.1103/PhysRevB.99.195150}

\bibitem{ogura} D. Ogura, H. Aoki, and K. Kuroki, {Phys. Rev. B {\bf{96}}, 184513 (2017).}
\url{https://doi.org/10.1103/PhysRevB.96.184513}

\bibitem{chamber}B. L. Chamberland, {Solid State Commun. {\bf{5}}, 663 (1967).}
\url{https://doi.org/10.1016/0038-1098(67)90088-9}

\bibitem{Zhou} J.-S. Zhou, C.-Q. Jin, Y.-W. Long, L.-X. Yang, and J. B. Goodenough, {Phys. Rev. Lett. {\bf{96}}, 046408 (2006).}
\url{https://doi.org/10.1103/PhysRevLett.96.046408}


\bibitem{Gupta} K. Gupta, P. Mahadevan, P. Mavropoulos, and M. Le\v{z}ai\'{c}, {Phys. Rev. Lett. {\bf{111}}, 077601 (2013).}
\url{https://doi.org/10.1103/PhysRevLett.111.077601}


\bibitem{sakurai} T. Yamauchi, T. Shimazu, D. Nishio-Hamane, and H. Sakurai, {Phys. Rev. Lett. {\bf{123}}, 156601 (2019).}
\url{https://doi.org/10.1103/PhysRevLett.123.156601}
\bibitem{weng} H. Weng, Y. Kawazoe, X. Wan, and J. Dong, {Phys. Rev. B {\bf{74}}, 205112 (2006).}
\url{https://doi.org/10.1103/PhysRevB.74.205112}

\bibitem{nozaku} J. Sugiyama, H. Nozaki, I. Umegaki, W. Higemoto, E. J. Ansaldo, J. H. Brewer, H. Sakurai, T-H Kao, H-D Yang, and M. M\r{a}nsson, 
{J. Phys.: Conf. Ser. {\bf{551}}, 012011 (2014).}
\url{https://doi.org/10.1088/1742-6596/551/1/012011}


\bibitem{takashi} T. Ishikawa, T. Toriyama, T. Konishi, H. Sakurai, and Y. Ohta, {J. Phys. Soc. Jpn. {\bf{86}}, 033701 (2017).}
\url{https://doi.org/10.7566/JPSJ.86.033701}


\bibitem{Hsakurai} H. Sakurai, {J. Phys. Soc. Jpn. {\bf{83}}, 123701 (2014).}
\url{https://doi.org/10.7566/JPSJ.83.123701} 


\bibitem{justin} J. Jeanneau, P. Toulemonde, G. Remenyi, A. Sulpice, C. V. Colin, V. Nassif, E. Suard, F. Gay, R. Weht, and M. N\'{u}\~{n}ez-Regueiro,
{EPL, {\bf{127}}, 27002 (2019).}
\url{https://doi.org/10.1209/0295-5075/127/27002}



\bibitem{yamaguchi}R. Takahashi, T. Yamaguchi, K. Sugimoto, T. Yamauchi, H. Sakurai, and Y. Ohta, {JPS Conf. Proc. {\bf{30}}, 011026 (2020).}
\url{https://doi.org/10.7566/JPSCP.30.011026}


\bibitem{zhu} Z. H. Zhu, W. Hu, C. A. Occhialini, J. Li, J. Pelliciari, C. S. Nelson, M. R. Norman, Q. Si, R. Comin, {arXiv:1906.04194 (2019).}
\url{https://arxiv.org/abs/1906.04194}


\bibitem{white}S. R. White,  {Phys. Rev. Lett. {\bf{69}}, 2863 (1992).} 
	\url{https://doi.org/10.1103/PhysRevLett.69.2863}

\bibitem{hundiron} For iron superconductors $J_{H}/U=0.25$ is often used. See Ref.~\cite{pdai} and 
Q. Luo {\it et al.}, Phys. Rev. B {\bf 82}, 104508 (2010).
\url{https://journals.aps.org/prb/abstract/10.1103/PhysRevB.82.104508}

\bibitem{Kresse:Prb} G. Kresse and J. Hafner, Phys. Rev. B \textbf{47}, 558(R) (1993).
	\url{https://doi.org/10.1103/PhysRevB.47.558}	
\bibitem{Kresse:Prb96} G.~Kresse and J.~Furthm\"{u}ller, Phys. Rev. B \textbf{54}, 11169 (1996). \url{https://doi.org/10.1103/PhysRevB.54.11169}
\bibitem{Blochl:Prb} P. E. Bl\"{o}chl, Phys. Rev. B \textbf{50}, 17953 (1994).
	\url{https://doi.org/10.1103/PhysRevB.50.17953}
\bibitem{Perdew:Prl} J. P. Perdew, K. Burke, and M. Ernzerhof, Phys. Rev. Lett. \textbf{77}, 3865 (1996). \url{https://doi.org/10.1103/PhysRevLett.77.3865}
\bibitem{Mostofi:cpc} A. A. Mostofi, J. R. Yates, Y. S. Lee, I. Souza, D. Vanderbilt, and N. Marzari, Comput. Phys. Commun. \textbf{178}, 685 (2007). \url{https://doi.org/10.1016/j.cpc.2007.11.016} 
\bibitem{Dudarev:prb} S. L. Dudarev, G. A. Botton, S. Y. Savrasov, C. J. Humphreys, and A. P. Sutton, Phys. Rev. B \textbf{57}, 1505 (1998).\url{https://doi.org/10.1103/PhysRevB.57.1505}

\bibitem{chuang}T.-M. Chuang, M. P. Allan, Jinho Lee, Yang Xie, Ni Ni, S. L. Bud’ko, G. S. Boebinger, P. C. Canfield, and J. C. Davis, {Science 327, {\bf{181}} (2010).}
\url{https://doi.org/10.1126/science.1181083 }

\bibitem{Luo}Q. Luo, A. Nicholson, J. Rinc\'on, S. Liang, J. Riera, G. Alvarez, 
	L. Wang, W. Ku, G. D. Samolyuk, A. Moreo, and E. Dagotto
	{Phys. Rev. B {\bf{87}}, 024404  (2013).}
	\url{https://doi.org/10.1103/PhysRevB.87.024404}

\bibitem{gonzalo}G. Alvarez, {Comput. Phys. Commun.{\bf{180}}, 1572 (2009).}
	\url{https://doi.org/10.1016/j.cpc.2009.02.016}


\bibitem{broyedn} D. D. Johnson, {Phys. Rev. B {\bf{38}}, 12807 (1988).}
	\url{https://doi.org/10.1103/PhysRevB.38.12807}

\bibitem{Matsuno:Prl} J. Matsuno, Y. Okimoto, M. Kawasaki, and Y. Tokura, Phys. Rev. Lett. \textbf{95}, 176404 (2005). \url{https://doi.org/10.1103/PhysRevLett.95.176404}

\bibitem{ladder1} E. Dagotto, J. Riera, and D. Scalapino, {Phys. Rev. B {\bf 45}, 5744(R) (1992)}.
\url{https://journals.aps.org/prb/abstract/10.1103/PhysRevB.45.5744}

\bibitem{ladder2} E. Dagotto and T. M. Rice, {Science 271, {\bf 618} (1996).}
\url{https://science.sciencemag.org/content/271/5249/618}

\bibitem{ladder3} M. Uehara, T. Nagata, J. Akimitsu, H. Takahashi, N. Mori, and K. Kinoshita
	, J. Phys. Soc. Jpn. {\bf 65}, 2764 (1996).
\url{https://doi.org/10.1143/JPSJ.65.2764}

\bibitem{ladder4} H. Takahashi, A. Sugimoto, Y. Nambu, T. Yamauchi, Y. Hirata, T. Kawakami, M. Avdeev, K. Matsubayashi, 
F. Du, C. Kawashima, H. Soeda, S. Nakano, Y. Uwatoko, Y. Ueda, T. J. Sato, and K. Ohgushi, Nat. Mater. {\bf 14}, 1008 (2015).
\url{https://www.nature.com/articles/nmat4351}

\bibitem{ladder5} N. D. Patel,  A. Nocera, G. Alvarez, R. Arita, A. Moreo, and E. Dagotto
	, Phys. Rev. B {\bf 94}, 075119 (2016).
\url{https://doi.org/10.1103/PhysRevB.94.075119}

\bibitem{ladder6} N. D. Patel, A. Nocera, G. Alvarez, A. Moreo, S. Johnston, and E. Dagotto, 
Commun. Phys. {\bf 2}, 64 (2019).
\url{https://www.nature.com/articles/s42005-019-0155-3}

\bibitem{orbmanga1} T. Hotta, M. Moraghebi, A. Feiguin, A. Moreo, S. Yunoki, and E. Dagotto
Phys. Rev. Lett. {\bf 90}, 247203 (2003).\url{https://doi.org/10.1103/PhysRevLett.90.247203}

\bibitem{orbmanga2} T. Hotta, S. Yunoki, M. Mayr, and E. Dagotto
Phys. Rev. B {\bf 60}, R15009(R) (1999).
\url{https://doi.org/10.1103/PhysRevB.60.R15009}

\bibitem{orbruthe} T. Hotta and E. Dagotto, Phys. Rev. Lett. {\bf 88}, 017201 (2001)
\url{https://doi.org/10.1103/PhysRevLett.88.017201}.

\bibitem{strainmanga} S. S. Hong, M. Gu, M. Verma, V. Harbola, B. Y. Wang, Di Lu, A. Vailionis, 
Y. Hikita, R. Pentcheva, J. M. Rondinelli, H. Y. Hwang, Science {\bf 368}, 71 (2020).
\url{https://science.sciencemag.org/content/368/6486/71}

\bibitem{cengiz} C. Sen and E. Dagotto, Phys. Rev. B {\bf 102}, 035126 (2020).
\url{https://doi.org/10.1103/PhysRevB.102.035126}



\bibitem{block1} J. Herbrych, J. Heverhagen, N. D. Patel, G. Alvarez, M. Daghofer, A. Moreo, and E. Dagotto, 
Phys. Rev. Lett. {\bf 123}, 027203 (2019).
\url{https://journals.aps.org/prl/abstract/10.1103/PhysRevLett.123.027203}

\bibitem{block2} N. Kaushal, A. Nocera, G. Alvarez, A. Moreo, and E. Dagotto, 
Phys. Rev. B {\bf 99}, 155115 (2019).
\url{https://doi.org/10.1103/PhysRevB.99.155115}

\bibitem{block3} J. Herbrych, G. Alvarez, A. Moreo, and E. Dagotto, Phys. Rev. B {\bf 102}, 115134 (2020).
\url{https://doi.org/10.1103/PhysRevB.102.115134}

\bibitem{block4} B. Pandey, L-F. Lin, R. Soni, N. Kaushal, J. Herbrych, G. Alvarez, and E. Dagotto, 
Phys. Rev. B {\bf 102}, 035149 (2020).
\url{https://doi.org/10.1103/PhysRevB.102.035149}

\bibitem{spiral} J. Herbrych, J. Heverhagen, G. Alvarez, M. Daghofer, A. Moreo, and E. Dagotto, 
Proc. Natl. Acad. Sci. USA {\bf 117}, 16226 (2020).
\url{https://www.pnas.org/content/117/28/16226}

\bibitem{patelnpj} N. D. Patel, N. Kaushal, A. Nocera, G. Alvarez, and E. Dagotto,
npj Quantum Mater. {\bf 5}, 27 (2020). 
\url{https://www.nature.com/articles/s41535-020-0228-2}

\end{thebibliography}
\end{document}